\documentclass[12pt,preprint]{aastex}

\def\umv{\hbox{\it u--v\/}}
\def\umy{\hbox{\it u--y\/}}
\def\umj{\hbox{\it u--J\/}}
\def\vmb{\hbox{\it v--b\/}}

\def\vmy{\hbox{\it v--y\/}}
\def\bmy{\hbox{\it b--y\/}}
\def\bmh{\hbox{\it b--H\/}}
\def\bmv{\hbox{\it B--V\/}}
\def\hmk{\hbox{\it H--K\/}}
\def\feh{\hbox{\rm [Fe/H]}}
\def\mh{\hbox{\rm [M/H]}}
\def\zx{\hbox{\rm Z/X}}
\def\afe{\hbox{\rm [$\alpha$/Fe]}}
\newcommand{\strom}{\mbox{Str\"omgren~}}

\shorttitle{New calibrations of the \strom metallicity index} 
\shortauthors{Calamida et al.}

\begin{document}



\title{\strom photometry of Galactic Globular Clusters. I. New 
Calibrations of the metallicity index.\altaffilmark{1}}

\author{
A. Calamida\altaffilmark{2,3},
G. Bono\altaffilmark{2,4},
P. B. Stetson\altaffilmark{5,10,11},
L. M. Freyhammer\altaffilmark{6},
S. Cassisi\altaffilmark{7}
F. Grundahl\altaffilmark{8},
A. Pietrinferni\altaffilmark{7},
M. Hilker\altaffilmark{4},
F. Primas\altaffilmark{4},
T. Richtler\altaffilmark{9}
M. Romaniello\altaffilmark{4},
R. Buonanno\altaffilmark{3},
F. Caputo\altaffilmark{2},
M. Castellani\altaffilmark{2},
C. E. Corsi\altaffilmark{2},
I. Ferraro\altaffilmark{2},
G. Iannicola\altaffilmark{2},
L. Pulone\altaffilmark{2}
}

\altaffiltext{1}{
Based in part on observations collected with the 1.54m Danish telescope
operated at ESO (La Silla) and with the Nordic Optical Telescope (NOT) 
operated at La Palma (Spain).}

\altaffiltext{2}{
INAF-Osservatorio Astronomico di Roma, Via Frascati 33, 00040, Monte Porzio 
Catone, Italy; bono@mporzio.astro.it, caputo@mporzio.astro.it, corsi@mporzio.astro.it, 
ferraro@mporzio.astro.it, giacinto@mporzio.astro.it, pulone@mporzio.astro.it}

\altaffiltext{3}{
Universit\`a di Roma Tor Vergata, Via della Ricerca Scientifica 1,
00133 Rome, Italy; buonanno@mporzio.astro.it, calamida@mporzio.astro.it}

\altaffiltext{4}{European Southern Observatory, Karl-Schwarzschild-Str. 2,
D-85748 Garching bei Munchen, Germany; mhilker@eso.org, fprimas@eso.org, 
mromanie@eso.org}

\altaffiltext{5}{
Dominion Astrophysical Observatory, Herzberg Institute of Astrophysics,
National Research Council, 5071 West Saanich Road, Victoria, BC V9E~2E7,
Canada; Peter.Stetson@nrc-cnrc.gc.ca}

\altaffiltext{6}{Centre for Astrophysics, University of Central Lancashire, 
Preston PR1 2HE, UK; lmfreyhammer@uclan.ac.uk}

\altaffiltext{7}{INAF-Osservatorio Astronomico di Collurania, via M. Maggini, 
64100 Teramo, Italy; cassisi@oa-teramo.inaf.it, adriano@oa-teramo.inaf.it 
}

\altaffiltext{8}{Department of Physics and Astronomy, Aarhus University,
Ny Munkegade, 8000 Aarhus C, Denmark; fgj@phys.au.dk
}

\altaffiltext{9}{Universidad de Concepcion, Departamento de Fisica, Casilla 
106-C, Concepcion, Chile; tom@mobydick.cfm.udec.cl 
}

\altaffiltext{10}{Guest User, Canadian Astronomy Data Centre, which is operated
by the Herzberg Institute of Astrophysics, National Research Council of Canada.}

\altaffiltext{11}{Guest Investigator of the UK Astronomy Data Centre.}


\begin{abstract}

We present a new calibration of the Stroemgren metallicity index $m_1$ 
using red giant (RG) stars in a sample of Galactic globular clusters 
(GGCs: M92, M13, NGC$\,$1851, 47~Tuc) that cover a broad range in 
metallicity ($-2.2 \le \feh \le -0.7$), are marginally affected by 
reddening uncertainties ($E(\bmv) \le 0.04$) and for which accurate 
$u,v,b,y$ Stroemgren photometry is available to well below the turnoff 
region. The main difference between the new empirical 
metallicity--index--color (MIC) relations and similar relations 
available in the literature is that we have adopted the \umy\ and 
\vmy\ colors instead of the \bmy\ color. These colors
present a stronger sensitivity to effective temperature, and 
the MIC relations show a linear and well-defined slope. 
The net difference between photometric estimates and spectroscopic 
measurements, for RG stars in five GGCs: M71, NGC$\,$288, NGC$\,$362, 
NGC$\,$6397, NGC$\,$6752,  is $0.04\pm 0.03$ dex with $\sigma$ = 0.11 dex. 
We also apply the new MIC relations to a sample of field stars for which 
spectroscopic metallicity ($-2.4 \le \feh \le -0.5$), accurate 
\strom photometry, and reddening estimates (Anthony-Twarog 
\& Twarog 1994, 1998) are all available. We find that
the difference between photometric estimates and spectroscopic measurements
is on average $-0.14\pm 0.01$ dex, with $\sigma = 0.17$ dex.\\
We also provide two independent sets
of MIC relations based on evolutionary models that have been transformed into
the observational plane by adopting either semi-empirical or theoretical 
color-temperature relations (CTRs). We apply the semi-empirical 
$\alpha-$enhanced MIC relations to the nine GCs and find that the difference
between photometric estimates and spectroscopic measurements is 
$0.04\pm 0.03$ dex, with $\sigma = 0.10$ dex. A similar agreement is also 
found for the sample of field stars, and indeed the difference is 
$-0.09\pm 0.03$ dex, with $\sigma = 0.19$ dex. 
The difference between metallicity estimates based on theoretical scaled-solar 
and spectroscopic measurements $-0.11\pm 0.03$ dex, with $\sigma=0.14$ dex 
for the nine GGCs and $-0.24\pm 0.03$ dex, with $\sigma=0.15$ dex for the 
field stars. 
On the whole, current findings support the evidence that new \strom MIC relations 
provide metallicity estimates with an intrinsic accuracy better than 0.2 dex.  
\end{abstract}

\keywords{globular clusters: general --- globular clusters: individual 
(M13, M71, M92, NGC$\,$288, NGC$\,$362, NGC$\,$1851, NGC$\,$6397, NGC$\,$6752, 47~Tuc)
 --- stars: abundances --- stars: evolution
}


\section{Introduction}\label{introduction}

The intermediate-band \strom photometric system (\strom 1966; Crawford 1975;
Bond 1980; Schuster \& Nissen 1988, hereinafter SN88) 
presents several indisputable advantages when compared
with broad-band photometric systems such as the Johnson-Cousins-Glass system
(Johnson \& Morgan 1953; Cousins 1976; Bessell 2005, and references therein). 
The key advantages of \strom photometry for A- to G-type stars are: 
{\em i\/}) the ability to provide robust estimates of intrinsic stellar parameters 
such as the metal abundance (the $m_1=(\vmb)-(\bmy)$ index,  
Richter et al.\ 1999; Anthony-Twarog \& Twarog 2000, hereinafter ATT00; 
Hilker 2000, hereinafter H00; Hilker \& Richtler 2000, hereinafter HR00), the 
surface gravity (the $c_1=(\umv)-(\vmb)$ index), and the effective temperature 
(the $H\beta$ index, Nissen 1988; Olsen 1988; ATT00). 
The $H\beta$ index is marginally affected by reddening, and therefore 
can also be compared to a simple color such as \bmy\  to provide individual estimates 
of reddening corrections (Nissen \& Schuster 1991). The same outcome applies to 
the reddening free $[c_1]$ index, and indeed theoretical and empirical 
evidence ({\it e.g.}, Stetson 1991; Nissen 1994; Calamida et al.\ 2005) suggests that 
a simple color such as \umy\ compared to $[c_1]$ (which is a temperature index
for stars hotter than 8,500 K) provides a robust reddening index for blue horizontal branch stars.    
{\em ii\/}) The ($[c_1]$,\,\vmy) color-color  plane provides robust estimates of 
the age of Galactic Globular Clusters (GGCs), since it is completely independent 
of cluster distance and marginally affected by uncertainties in interstellar 
reddening corrections. Moreover, the region around the main-sequence turnoff (TO) presents 
a cuspy shape in this diagram (see Fig.\ 2 in Grundahl et al.\ 1998), and therefore its 
identification is more robust than in the typical Johnson-Cousins bands. 
In the latter photometric system the region around the TO presents a steep 
slope (Rosenberg et al.\ 2000).   
{\em iii\/}) The ($u$,\,\umy) Color-Magnitude Diagram (CMD) provides the opportunity 
to identify a {\em jump\/} among hot Horizontal Branch stars at 
$11,500 \lesssim T_{eff} \lesssim 12,000$ K 
caused by radiative levitation of metals (Grundahl et al.\ 1998, 1999).   
{\em iv\/}) Detailed empirical investigations characterized the \strom 
system not only for $A-G$ type dwarfs (Crawford 1975, 1979; Nissen 1988; 
Olsen 1988; Schuster \& Nissen 1989, hereinafter SN89; Nordstrom et al.\ 2004), 
but also for $G-K$ type giants (Bond 1980; Richtler 1989; Twarog \& 
Anthony-Twarog 1991; Grebel \& Richtler 1992; Anthony-Twarog \& Twarog 1994, 
hereinafter ATT94; H00).  
{\em v\/}) The use of the $m_1$ versus color plane can also be safely 
adopted to distinguish cluster and field stars (ATT00; Rey et al.\ 2004). 
{\em vi\/}) Accurate \strom photometry can also be adopted 
to constrain the ensemble properties of stellar populations in complex 
stellar systems like the Galactic bulge (Feltzing \& Gilmore 2000)
and disk (Haywood 2001).  
{\em vii\/}) \strom photometry has been recently adopted to investigate 
the membership and the metallicity distribution of Red Giant (RG) stars 
in the Local Group dwarf spheroidal galaxy Draco (Faria et al.\ 2007). 
Moreover, it has also been adopted to remove the 
degeneracy between age and metallicity in other stellar systems hosting 
simple stellar populations (GCs, elliptical galaxies), to investigate 
age and metallicity distributions of dwarf elliptical galaxies in 
the Coma and Fornax galaxy clusters (Rakos \& Schombert 2004, 2005).         

On the other hand, the \strom system presents two substantial 
drawbacks. {\em i\/}) the $u$ and $v$ bands have short effective  
wavelengths, namely $\lambda_{eff}\,=\,3450$ and $\lambda_{eff}\,=\,4110\,$\AA. 
As a consequence the ability to perform accurate photometry 
with current CCD detectors is hampered by their reduced sensitivity 
in this wavelength region. 
{\em ii\/}) The intrinsic accuracy of the stellar parameters,  
estimated using \strom indices, strongly depends on 
the accuracy of the absolute zero-point calibrations. This 
typically means an accuracy better than 0.03 mag. This limit 
could be easily accomplished in the era of photoelectric photometry,
but it is not trivial at all in the modern age of CCDs.

The use of \strom photometry was also hampered by the lack 
of accurate bolometric corrections (BCs) and color-temperature 
relations (CTRs) based on recent and homogeneous sets of atmosphere 
models. This gap was partially filled by the new semi-empirical 
set of BCs and CTRs provided by Clem et al.\ (2004, hereinafter CVGB04) 
and by the new theoretical calibration of the $H\beta$ index provided 
by Castelli \& Kurucz (2006, hereinafter CK06). Moreover and even more 
importantly, current empirical calibrations of \strom metallicity indices are 
based either on field stars (ATT94) or on a mix of cluster and field stars (H00).  
However, empirical spectroscopic evidence suggests that 
field and cluster stars  present different heavy element abundance
patterns (Gratton, Sneden \& Carretta 2004). 
Moreover, the occurrence of CN and/or CH rich stars in GGCs (Anthony-Twarog,
Twarog, \& Craig 1995; Grundahl, Stetson, \& Andersen 2002) along the RG (HR00),
the subgiant branch, and the main sequence (Stanford et al.\ 2004; Kayser et
al.\ 2006) also suggests the opportunity for an independent
calibration\footnote{The referee noted that the calibration of the metallicity
index based on a mixing of field disk stars and open cluster stars does not show
any drawback (Twarog et al.\ 1997)} of the \strom metallicity index based
only on cluster stars as originally suggested by Richtler (1989).

To fill this gap we plan to provide new empirical, semi-empirical,
and theoretical calibrations of the $m_1$ metallicity index
using cluster stars, and new sets of semi-empirical and theoretical
transformations. This is the first paper of a series devoted to \strom 
photometry of GGCs. The structure of the current paper is as follows. 
In \S 2 we discuss in detail the photometric catalogs we adopted for 
the new empirical calibration
and for validating current metallicity-index--color (MIC) relations.
Section 3 deals with the selection criteria adopted to select the GCs for the
calibration together with the optical-NIR two-color planes and the proper-motion
selection adopted to identify candidate field and cluster RG stars. In \S 4 we 
discuss the approach adopted to calibrate the \strom metallicity index, while 
in \S 5 we present the different tests we performed to validate the current empirical
calibrations and the comparison between photometric estimates and spectroscopic
measurements of iron abundances. Section 6 deals with the calibration of both
semi-empirical and theoretical MIC relations.  In this section we also discuss
the validation of the new relations and the comparison with spectroscopic
abundances and with other calibrations of the \strom MIC relations available in
the literature. We summarize the results and briefly discuss further
improvements and applications of the new MIC relations in \S 7.

\section{Observations and data reduction}\label{observations and data reduction}

The photometric catalogs of globular clusters adopted in this investigation 
were collected with the 2.56m Nordic Optical Telescope (NOT) on La Palma  
and with the 1.54m Danish Telescope on La Silla (ESO), using the $uvby$ filter 
sets available there (see Table~1 for a log of the observations).  
Data secured with the NOT were collected during three observing runs in 1995, 
1997, and in 1998. Stars from the lists of Olsen (1983, 1984) and 
SN88 were observed on two nights 
in 1995 and four nights in 1998 under photometric conditions, to derive 
the transformation between the instrumental magnitudes and the standard 
system. The data for M13 have already been described in Grundahl et al.\ (1998), 
while those for M92 (NGC~6341) were collected with a thinned AR coated 
$2048 \times 2048$ pixel CCD camera on the HiRAC instrument, with 0\farcs11 per pixel, thus 
covering a sky area of approximately $3.75\times3.75$ arcmin$^2$. 
Most of the observations were collected using tip/tilt correction,
and the seeing FWHM of the entire set of images ranges from $\sim$0\farcs45 to 
$\sim$1\farcs0. There was no significant variation 
of the point spread function (PSF) over the field of view. We observed 
two overlapping fields in M92, with one field on the cluster 
center to ensure a large sample of HB and red-giant branch (RGB) stars.
Data for M71 were collected between June 26 and July 2, 1995, and we 
observed a field 2${}^\prime$ north of the cluster center (for more details 
see Grundahl et al.\ 2002). 

The images from the 1.54m Danish Telescope were acquired during 
two observing runs in May and in October 1997. For both runs we used 
the Danish Faint Object Spectrograph and Camera (DFOSC) equipped with 
a thinned, AR coated  2048 $\times$ 2048 pixel CCD camera. The field 
of view covered by these data is approximately 11 arcmin across. 
During the October observing run data were collected for NGC$\,$104, NGC$\,$288, 
NGC$\,$1851, NGC$\,$362, and NGC$\,$6752, and during the run in May 
for NGC$\,$6397 (see Table~1). 
The selected clusters were observed on several photometric nights, and 
approximately 150 different standard stars from the list collected by 
Olsen (1983, 1984) and by SN88 were also observed. These images were 
secured during seeing conditions ranging from 1\farcs3 to 2\farcs2. 
Flat fields were obtained on each night during evening and 
morning twilight. Photometry for the defocused standard stars was 
derived from large-aperture photometry.  

The photometry of the cluster and standard frames was performed with
DAOPHOT$\,${\footnotesize IV}/ALLFRAME and DAOGROW (Stetson 1987, 1991, 1994).
Based on the frame--to--frame scatter 
for the bright stars in the clusters with calibrated photometry we 
estimate that the errors in the photometric zero points are below 
0.02~mag for the observations from NOT, and less than 0.03~mag for 
the data from the 1.54m Danish Telescope. 
The reader interested in more details concerning the observations, 
data reduction and calibration procedures is referred to  Grundahl et al.\ (1999) 
and Grundahl, Stetson \& Andersen (2002).  The final calibrated 
cluster catalogs include $\sim$ 15,000--30,000 stars.
The \strom catalogs adopted in this investigation can be retrieved from the
following URL: {\tt http://www.mporzio.astro.it/spress/stroemgren.php}.

  \section{Globular cluster selection}\label{clu}

In order to calibrate the metallicity index $m_1$ we selected four globular
clusters, namely M92, M13, NGC$\,$1851, NGC$\,$104, that cover a broad range in 
metallicity ($-2.2<\feh<-0.7$), are marginally affected by reddening 
($E(\bmv) \le 0.04$), and for which accurate \strom photometry is available to
well below the turnoff region (Grundahl et al.\ 1998; 
Grundahl et al.\ 1999, Grundahl, Stetson \& Andersen 2002).
We performed several tests by including among the calibrating clusters 
other GCs that also have low reddening values (NGC$\,$288, NGC$\,$362, NGC$\,$6752), 
but the intrinsic accuracy of the calibration did not improve, since their iron 
abundances are very similar either to M13 or to NGC$\,$1851. 
The metallicities and the reddening values for these clusters 
are listed in Table~2. Empirical evidence suggests that the $m_1$ versus color 
relation of RG stars presents a linear trend and a good sensitivity 
to iron abundance 
(Bond 1980; Richtler 1989; Twarog \& Anthony-Twarog 1991; Grebel \& Richtler 1992; 
ATT94; H00). Therefore, we selected cluster stars from 
the tip to the base of the RGB with a photometric accuracy 
$\sigma_{u,v}\le$ 0.03 mag and $\sigma_{b,y}\le$ 0.02 mag for 
each cluster in our sample.

However, in order to avoid subtle systematic uncertainties in the empirical
calibrations, actual cluster RG stars need to be distinguished from 
contaminating field stars.
To accomplish this goal we decided to use optical-NIR color--color planes
to split cluster from field stars. In particular, we cross-identified stars
in common with our \strom catalogs and the
Near-Infrared (NIR) Two Micron All Sky Survey (2MASS) catalog
(Skrutskie et al.\ 2006\footnote{See also
http://www.ipac.caltech.edu/2mass/releases}).
Moreover, we also re-identified a subsample of our optical catalog in the
second US Naval Observatory CCD Astrograph proper motion catalog
(UCAC2, Zacharias et al.\ 2004).
In both cases, the cross identification was performed following these steps:
{\em i\/}) IRAF's IMMATCH package was used to establish a preliminary
spatial transformation from the \strom catalog's CCD coordinates to the
reference catalog's Equatorial (J2000.0) system for a subsample of matched
stars; {\em ii\/}) the full \strom catalog was transformed onto and matched
with the reference catalog, on the basis of radial distance and 
initially also on apparent stellar brightness; {\em iii\/}) the previous 
steps were reiterated 2--3 times until the transformation permitted
a near-complete matching, and then, {\em iv\/}) the final, matched sample
of common stars was obtained by rejecting entries separated by more than,
typically, 0\farcs8--1\farcs3.

In certain cases the stellar magnitudes were used also to reject
mismatches while taking into account possible ranges of color differences
in the available bands. Note that our \strom photometry mostly covers 
relatively small and off-center fields of the analyzed clusters.
The stated  UCAC2 proper-motion errors are about 1--3 mas yr$^{-1}$
for stars to 12th magnitude and 4--7 mas yr$^{-1}$ for fainter stars
to 16th magnitude, while the precision of the positions is 15--70 mas, 
depending on magnitude, with
estimated systematic errors of 10 mas or below.  The  2MASS catalog has  
limiting magnitudes for the $J$, $H$, and $Ks-$bands of about 15.8, 15.1, and 
14.3 mag, respectively, while the astrometric accuracy is of the order of 100 mas.

\begin{figure}[ht!]
\begin{center}
\label{fig1}
\includegraphics[height=0.45\textheight,width=0.60\textwidth]{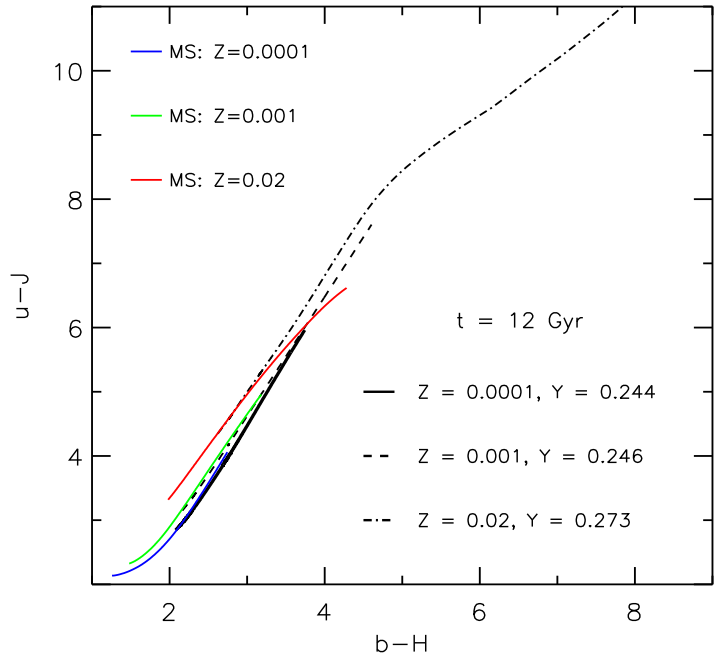}
\caption{Optical-Near-Infrared color-color plane for isochrones at fixed 
cluster age and different chemical compositions (see labeled values). 
Evolutionary phases included between the base (hot end) and the 
tip (cool end) of the RGB are shown in black. Evolutionary phases 
included between the central H-burning and the turnoff point  
are shown in colors.
Evolutionary tracks have been computed assuming scaled-solar 
compositions (Pietrinferni et al.\ 2004). Theoretical predictions 
have been transformed into the observational plane by adopting 
atmosphere models based on the same scaled-solar abundances 
adopted in the evolutionary computations.}
\end{center}
\end{figure}

Finally, we selected for each cluster only the RG stars with at least 
three NIR measurements ($J,H,K$) and all four \strom magnitudes. We found 
that ($\umj,\,\bmh)$ is the best optical-NIR color-color plane to properly 
identify field and cluster stars. Fig.~1 shows three isochrones at fixed 
cluster age ($t = 12$ Gyr) and different chemical compositions in this plane. 
The evolutionary models and the atmosphere models adopted to transform 
theoretical predictions into the observational plane have been constructed 
adopting a scaled-solar abundance mixture. The evolutionary phases plotted 
in this figure range from the base (hot end) to the tip (cool end) of the RGB. 
The systematic drift, at fixed $\bmh$, toward redder colors when moving from 
metal-poor to metal-rich structures is clear. It is noteworthy that a difference  
of 1,200 K (Z=0.0001) and of 2,000 K (Z=0.02) between the base and the 
tip of the RGB are covered by $\sim$ three and by more than seven 
magnitudes, respectively. Together with the RG evolutionary phases Fig.~1
also shows the central H-burning phases up to the turnoff point for the 
same metal abundances and cluster ages (solid colored lines). A partial 
degeneracy between metal-poor and metal-intermediate dwarfs and RG stars 
takes place only in the hot corner of this color-color plane, while at 
solar chemical composition it covers a broader color range. However, 
typical halo and thin disk field populations possess iron abundances 
systematically more metal-poor than the solar value (Castellani et al.\ 
2002).  
We did not use the $K$-band photometry because it is less accurate in the 
faint magnitude limit when compared with $J$- and $H$-band magnitudes. 
A similar procedure but based on Johnson-Cousins optical magnitudes and NIR 
magnitudes was adopted by Castellani et al.\ (2007) to identify probable 
cluster and field stars in $\omega$ Centauri. The interested reader is 
referred to this paper for a detailed discussion concerning the approach 
adopted to select field and cluster stars.   

\begin{figure}[ht!]
\begin{center}
\label{fig2}
\includegraphics[height=0.5\textheight,width=0.6\textwidth]{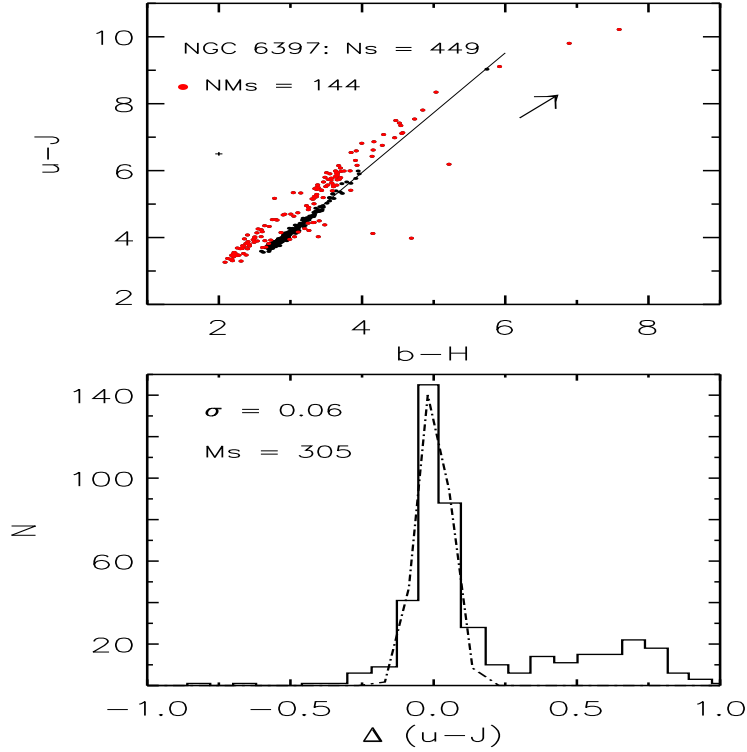}
\caption{Top -- RG stars of NGC$\,$6397 plotted in the optical--NIR color--color 
$(\umj,\,\bmh)$ plane. Red dots are candidate field stars (Non-Member stars, NMs = 144). 
The solid line is the fitted cluster fiducial sequence. The arrow shows the reddening 
vector. Bottom -- Distribution of the difference between the \umj\ color of 
individual stars and the \umj\ color of the fiducial cluster sequence. 
The dash--dotted line displays the Gaussian function that fits the main peak in the 
color difference distribution. Objects with $\Delta(\umj) \le 3\times \sigma_{\umj}$ 
were considered candidate cluster RG stars (Member stars, Ms = 305).}
\end{center}
\end{figure}

Fig.~2 shows NGC$\,$6397 RG stars (449 of them) plotted on the aforementioned color--color 
plane. Metal-poor cluster stars form a narrow sequence ranging
from $\bmh \sim 2.5$ to $\bmh \sim 4.0$, while more metal-rich 
candidate field stars are distributed along a separate sequence,  
systematically redder in \umj\ at fixed \bmh\ color. 
The referee noted that in this color-color plane the different stellar populations
present a slope very similar to the slope of the reddening vector. This means that
the selection between cluster and field stars is minimally affected by a difference
in reddening.
The objects with $\bmh >$ 5.0 
are probably metal-rich field star candidates. Once we identified the 
fiducial cluster sequence in the $(\umj,\,\bmh)$ plane we performed a 
linear fit, $\umj = \alpha + \beta (\bmh)$, for the candidate cluster 
stars. Then, we estimated the difference in \umj\ color between individual 
RG stars and the fiducial line at the same \bmh\ color. The bottom panel of 
Fig.~2 shows the distribution of the color excess $\Delta (\umj)$ for 
the entire sample. We fitted the distribution with a Gaussian function and we 
considered only those stars with $\Delta(\umj) \le 3\times \sigma_{\umj}$
as {\em bona fide\/} cluster RG stars. The red dots in the top 
panel of Fig.~2 mark the candidate field stars after this 
selection. The original sample was thus reduced by roughly 40\%. 
Subsequently, we also applied a selection by proper motion. 
In particular, we considered as cluster members those stars with proper 
motions smaller than $35~mas/yr$, and 
$\vert\arctan\big(\frac{P_{RA}}{P_{DEC}}\big)\vert \le$ 1.0. This additional 
selection decreased the sample of candidate RG members by less than 
10\%. 

\begin{figure}[ht!]
\begin{center}
\label{fig3}
\includegraphics[height=0.45\textheight,width=1.0\textwidth]{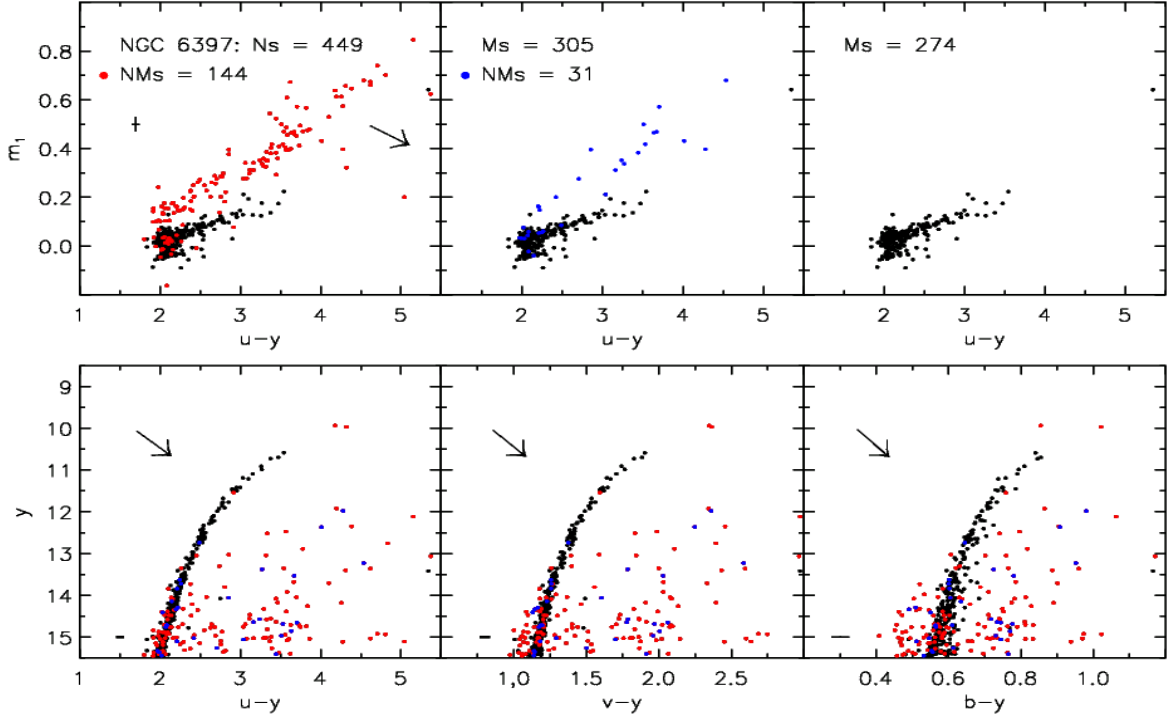}
\caption{Top: RG stars in NGC$\,$6397 plotted in the ($m_1,\ \umy)$ plane (left). 
Candidate cluster stars were selected according to the optical-NIR color-color 
plane ($\Delta (\umj) \le 3.0\times \sigma_{\umj}$) and to the proper motion 
velocity 
($\le 35\, mas/yr$ and $\vert\arctan\big(\frac{P_{RA}}{P_{DEC}}\big)\vert \le$ 1.0).
The red dots mark probable field stars selected according 
to optical-NIR colors (NMs = 144). The error bars account for uncertainties in 
intrinsic photometric errors. The arrow shows the reddening vector.  
The black dots in the middle panel display the candidate cluster RG stars 
according to the optical-NIR color-color plane selection (Ms = 305). Blue dots 
mark probable field stars according to the PM selection (NMs = 31). The right 
panel shows candidate cluster RG stars (Ms = 274) according to the two selection 
criteria. 
Bottom: \strom Color-Magnitude Diagrams $y$,\umy (left), $y$,\vmy
(middle), and $y$,\bmy (right) for cluster and field star candidates. 
}
\end{center}
\end{figure}

In order to verify the reliability of the selection procedure that we devised 
to distinguish probable cluster and field stars, the top panels of Fig.~3 show
from left to right the distribution in the ($m_1$,\,\umy) color--color plane of the original 
RG sample (449 stars), of the candidate cluster RGs after the selection in 
the optical--NIR color-color plane (305), and of the candidate RGs after the
selection by proper motion (274). A few interesting features of the 
($m_1$,\,\umy)  plane must be discussed in detail: 
{\em i\/})~we adopted the \umy\ color as a temperature index. The 
main advantage of this color over \bmy\ is the stronger temperature sensitivity.
The RG stars in NGC$\,$6397 cover more 
than two magnitudes in \umy\ while the same objects cover only 0.5
mag in \bmy. Obviously, the reddening correction for the \umy\ color is 
larger than for the \bmy\ color, but the reddening toward the selected 
calibrating GCs is relatively well known and they are not affected, 
according to current empirical evidence, by differential reddening.    
{\em ii\/})~Data plotted in the top left panel of Fig.~3 show a double 
stellar sequence. The sequence that attains larger $m_1$ at fixed \umy\ 
values almost completely disappears after the selection in the color--color 
plane (see the middle panel). The Proper Motion (PM) selection decreases 
by roughly the 10\% the sample of candidate cluster stars. It was originally 
suggested by Bell \& Gustafsson (1978) and more recently by ATT94, H00, and 
by Grundahl et al.\ (2002) that stars with large $m_1$ values present 
an over-abundance of carbon and/or nitrogen, i.e. they might be CN- 
and/or CH-rich stars.   
As a matter of fact, two strong cyanogen (CN) molecular absorption bands 
are located at $\lambda=4142$ and $\lambda=4215$ \AA, i.e. very close to 
the effective wavelength of the $v$ filter ($\lambda_{eff}=4110$,$\;\; 
\Delta \lambda=190$ \AA). Moreover, the strong CH molecular band 
located in the Fraunhofer's $G-$band ($\lambda=4300$ \AA) might affect 
both the $v$ and the $b$ magnitude. It is noteworthy that the molecular 
NH band  at $\lambda=3360$ \AA, and the two CN bands at $\lambda=3590$ 
and $\lambda=3883$ \AA~ might affect the $u$ ($\lambda_{eff}=3450$,$\;\;
\Delta \lambda=300$ \AA) magnitude (see e.g. Smith 1987).    
However, current evidence based on optical-NIR color-color and on proper 
motion selections suggests that the stars with larger $m_1$ values 
in NGC$\,$6397 could be field stars. Note that a mild correlation between a 
NIR (\hmk) color-excess and the strength of the CN band at $\lambda=4215$ \AA 
--based on the $C_m$ index of the DDO photometric system-- was detected by 
Smith (1988) in a sample of Population {\small I} CN-rich field giants. However, 
we are not aware of any empirical evidence suggesting that CN-rich stars 
in GCs also show a NIR color-excess.  

\begin{figure}[ht!]
\begin{center}
\label{fig4}
\includegraphics[height=0.70\textheight,width=0.55\textwidth]{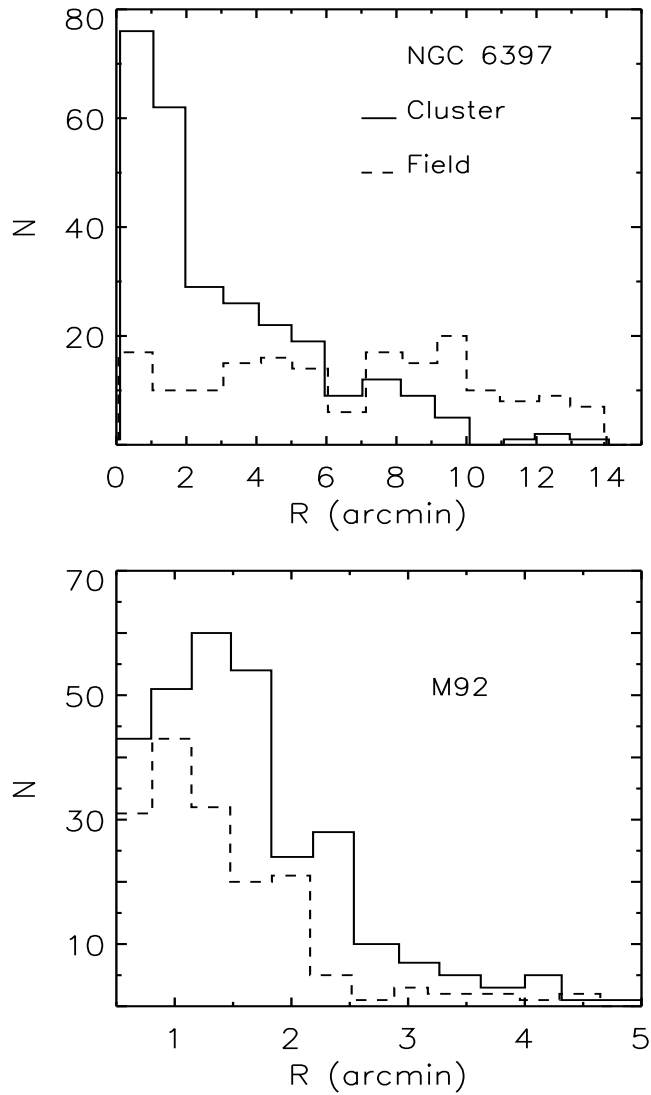}
\caption{
Top - Radial distribution of candidate cluster RG stars (solid line) and
of probable field stars (dashed line) for the GC NGC~6397. The latter sample
includes stars selected according to the color-color plane and to the proper
motions. R is the distance from the center of the cluster in arcminutes.
Bottom - Same as the top, but for the GC M92.
}
\end{center}
\end{figure}

As a further check, the same RGs samples have been plotted in 
three different CMDs in the bottom panels of Fig.~3, namely 
($y$,\,\umy) (left) ($y$,\,\vmy) (middle), and 
($y$,\,\bmy) (right). A glance at the data plotted in this figure shows 
that the {\em bona fide\/} cluster RG stars (black dots) are distributed along 
a very narrow color-magnitude sequence. On the other hand, a large fraction of 
the sequence with large $m_1$ values (red dots) covers a broad range in both color 
and magnitude. It is noteworthy that several of these stars and a good fraction 
of the stars with large PM values possess magnitudes and colors that are 
very similar to candidate cluster RG stars. This finding further supports the use 
of the optical-NIR color-color plane to properly separate field and cluster stars. 
However, we cannot exclude the possibility that a fraction of current candidate non-members 
are cluster stars with peculiar spectra. 

As a final validation of the selection procedure, we compared the radial 
distributions of candidate cluster and field stars. The top panel of Fig.~4 
shows the two 
distributions and the flat distribution of field stars (dashed line) is quite 
evident when compared with the steeper and more centrally concentrated 
distribution of {\em bona fide\/} cluster stars (solid line). The mild decrease 
in the number of field stars in the outer reaches of the cluster suggests
that the color-color selection we applied is very conservative, and some real
cluster members might have been erroneously rejected. These objects deserve 
spectroscopic follow-up to determine whether their peculiar optical-NIR 
colors are caused either by a significantly different chemical composition 
or by the presence of secondary companions or both.  

The same approach was adopted for selecting the {\em bona fide\/} cluster 
RG stars of the other calibrating clusters. In particular,  Fig.~5 shows 
the selection applied to RGs in the metal-poor GC M92. For this 
cluster we considered only those stars with 
$\Delta {\umj} \le 1.5\times \sigma_{\umj}$ as candidate RG members.
Once again the original sample 
was reduced by approximately 40\% after the selection in the 
color-color plane was applied. 
%
\begin{figure}[!ht]
\begin{center}
\label{fig5}
\includegraphics[height=0.6\textheight,width=0.5\textwidth]{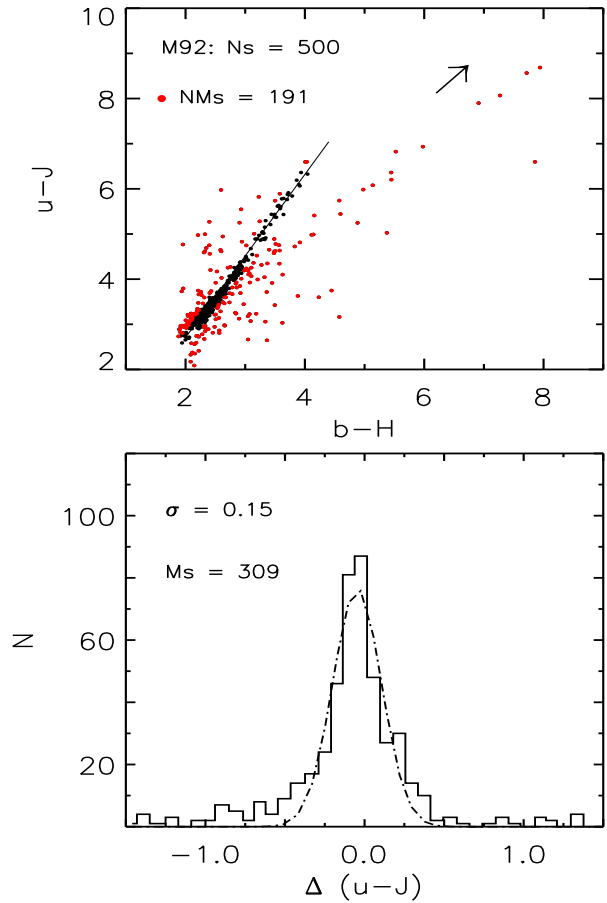}
\caption{Same as Fig.~2, but for RG stars in the GC M92. Only the objects 
with $\Delta (\umj) \le 1.5\times \sigma_{\umj}$ were considered candidate 
cluster RG stars.}
\end{center}
\end{figure}
\begin{figure}[!ht]
\begin{center}
\label{fig6}
\includegraphics[height=0.7\textheight,width=0.6\textwidth,angle=90]{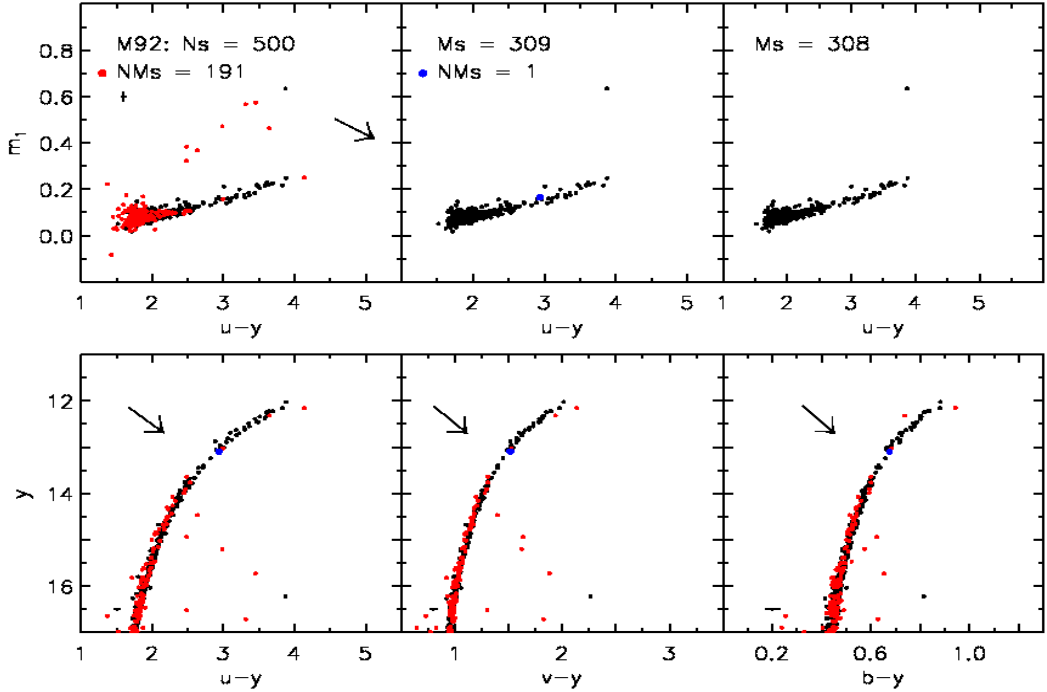}
\caption{Same as Fig.~3, but for RG stars in the GC M92. Only the 
RG stars with a membership probability $P_{c} \geq$ 90 were considered 
candidate cluster members (Cudworth 1976).}
\end{center}
\end{figure}

Proper motion measurements for this cluster are not available in the 
UCAC2 catalog, therefore we adopted the measurements provided by 
Cudworth (1976). 
We cross-identified our \strom catalog with the Cudworth catalog and we found 50 RG
stars in common. Among them three have a cluster membership probability $P_{c} <$ 90,
and two were removed according to the optical-NIR color-color selection.
The top panels of Fig.~6 show the M92 RG stars in the ($m_1$,\,\umy) plane 
before (left), after the optical-NIR color-color selection (middle), and
after the proper motion selection (right).
It is worth noting that the stars that attain large $m_1$ values, at 
fixed \umy\ color, disappear after the selection in the $(\umj,\,\bmh)$ plane. 
Moreover, almost all candidate field stars (red dots in Figs. 5 and 6)
present magnitudes and colors similar to candidate cluster RG stars (see 
bottom panels of Fig.~6). The same outcome applies to the RG stars in M13, 
and indeed after the selection in the optical-NIR color-color plane the 
cluster RG candidates occupy a narrow and well-defined sequence. 
Data plotted in the bottom panel of Fig.~4 show that the radial distribution
of probable field stars in M92 is quite flat in the external regions, but becomes
similar to the candidate cluster RG stars in the innermost regions. However,
as already mentioned above and by ATT00 the key point in current selections is
more to leave the probable nonmembers out than to keep the candidate members in.

The referee suggested that we comment on the different distribution of field stars
in the optical-NIR color-color plane between NGC~6397 and M92. In order to provide a
quantitative estimate we performed two simulations of the field star distribution
using the Galactic model developed by Cignoni et al. (2006, and references therein).
We find that the field across NGC~6397 consists of 19\% halo, 39\% thick disk, and
42\% thin disk stars (Castellani et al.\ 2001). On the other hand, the field across
M92 consists of 38\% halo, 36\% thick disk, and 26\% thin disk stars. The above
numbers indicate that the main difference in the M92 field is due to the substantial
decrease in the fraction of more metal-rich thin disk stars and in the increase in
the fraction of less metal-rich halo stars.

\begin{figure}[!ht]
\begin{center}
\label{fig7}
\includegraphics[height=0.6\textheight,width=0.5\textwidth]{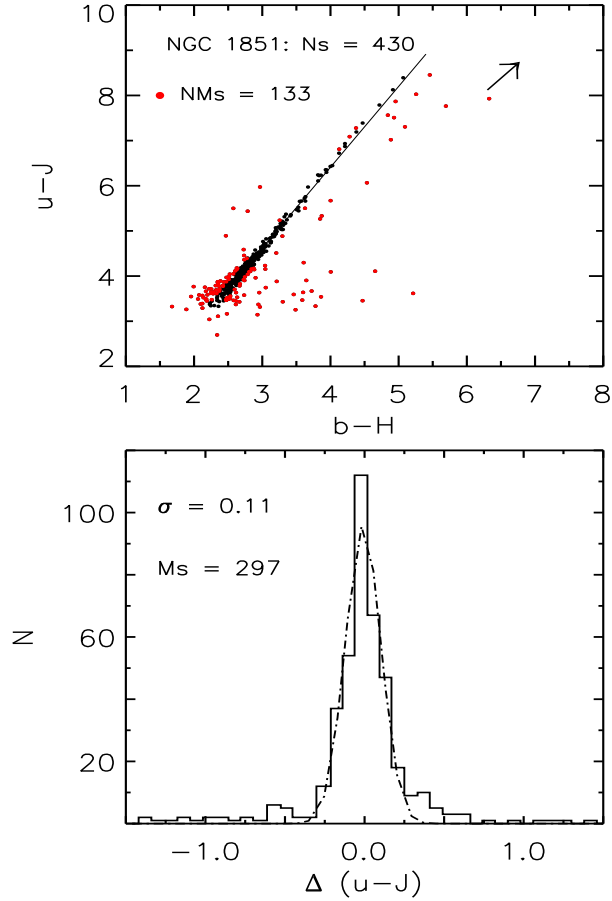}
\caption{Same as Fig.~2, but for RG stars in the GC NGC$\,$1851. Only the objects 
with $\Delta (\umj) \le 1.5\times \sigma_{\umj}$ were considered candidate 
cluster RG stars.}
\end{center}
\end{figure}

\begin{figure}[!ht]
\begin{center}
\label{fig8}
\includegraphics[height=0.75\textheight,width=0.65\textwidth,angle=90]{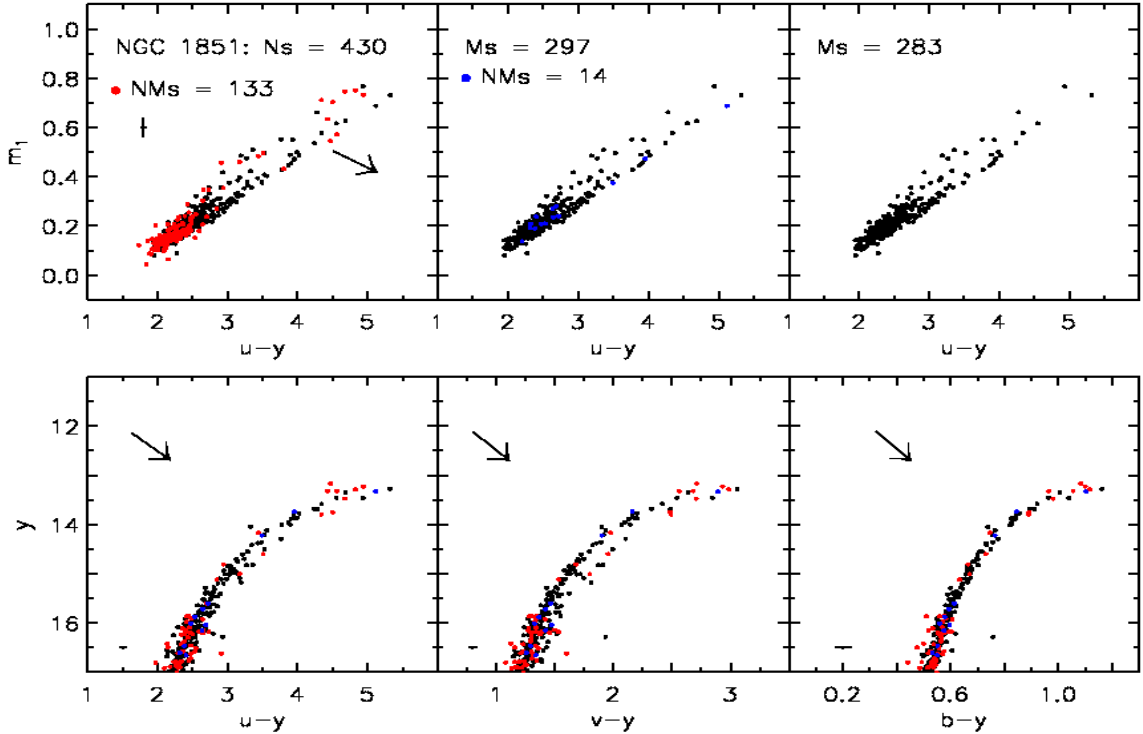}
\caption{Same as Fig.~3, but for RG stars in NGC$\,$1851. Note that the double 
sequence of candidate cluster RG stars with large $m_1$ values does not 
disappear after the selections in the optical-NIR color-color plane (top middle panel) 
and for proper motions (top right panel). The RG stars with a proper motion velocity 
$\le 7~mas/yr$ were considered candidate cluster members.}
\end{center}
\end{figure}

On the other hand, RG stars in NGC$\,$1851 and in NGC$\,$104 present, after the
selection in optical-NIR color--color plane (Figs. 7 and 9), and the selection for
proper motions (see top panels of Figs. 8 and 10), the evidence either of a double
sequence (NGC$\,$1851) or of a large spread in $m_1$ values (NGC$\,$104). It is
noteworthy that this spread is still present among candidate cluster stars and
that the fraction of stars redder than \umy\ = 2.8 and with large $m_1$ values
is $\approx 50$\% for RGs in NGC$\,$1851 and $\approx 60$\% for RGs in NGC$\,$104.

\begin{figure}[!ht]
\begin{center}
\label{fig9}
\includegraphics[height=0.6\textheight,width=0.5\textwidth]{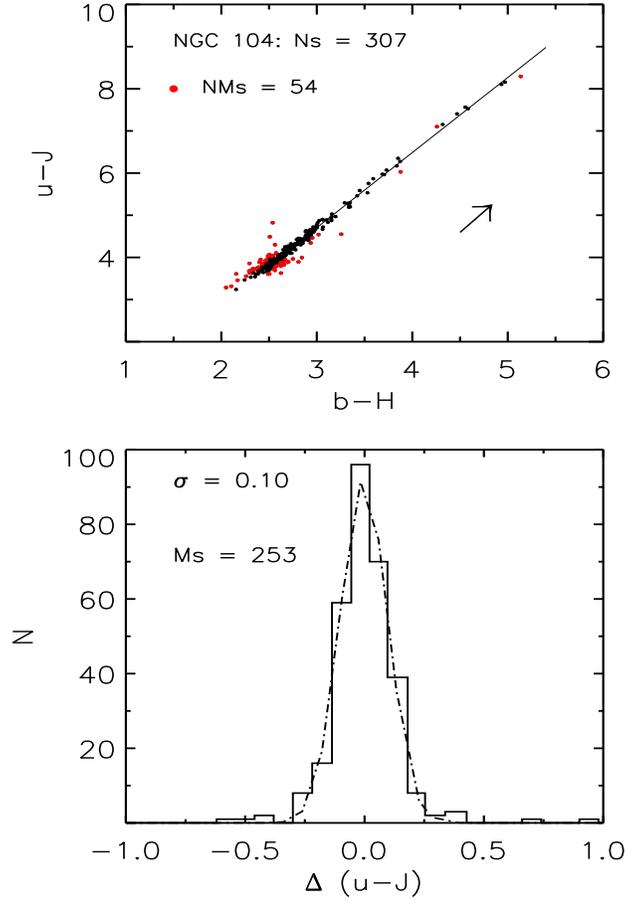}
\caption{Same as Fig.~2, but for RG stars in the GC NGC$\,$104. Only the objects 
with $\Delta (\umj) \le 1.5\times \sigma_{\umj}$ were considered candidate 
cluster RG stars.}
\end{center}
\end{figure}

\begin{figure}[!ht]
\begin{center}
\label{fig10}
\includegraphics[height=0.75\textheight,width=0.65\textwidth,angle=90]{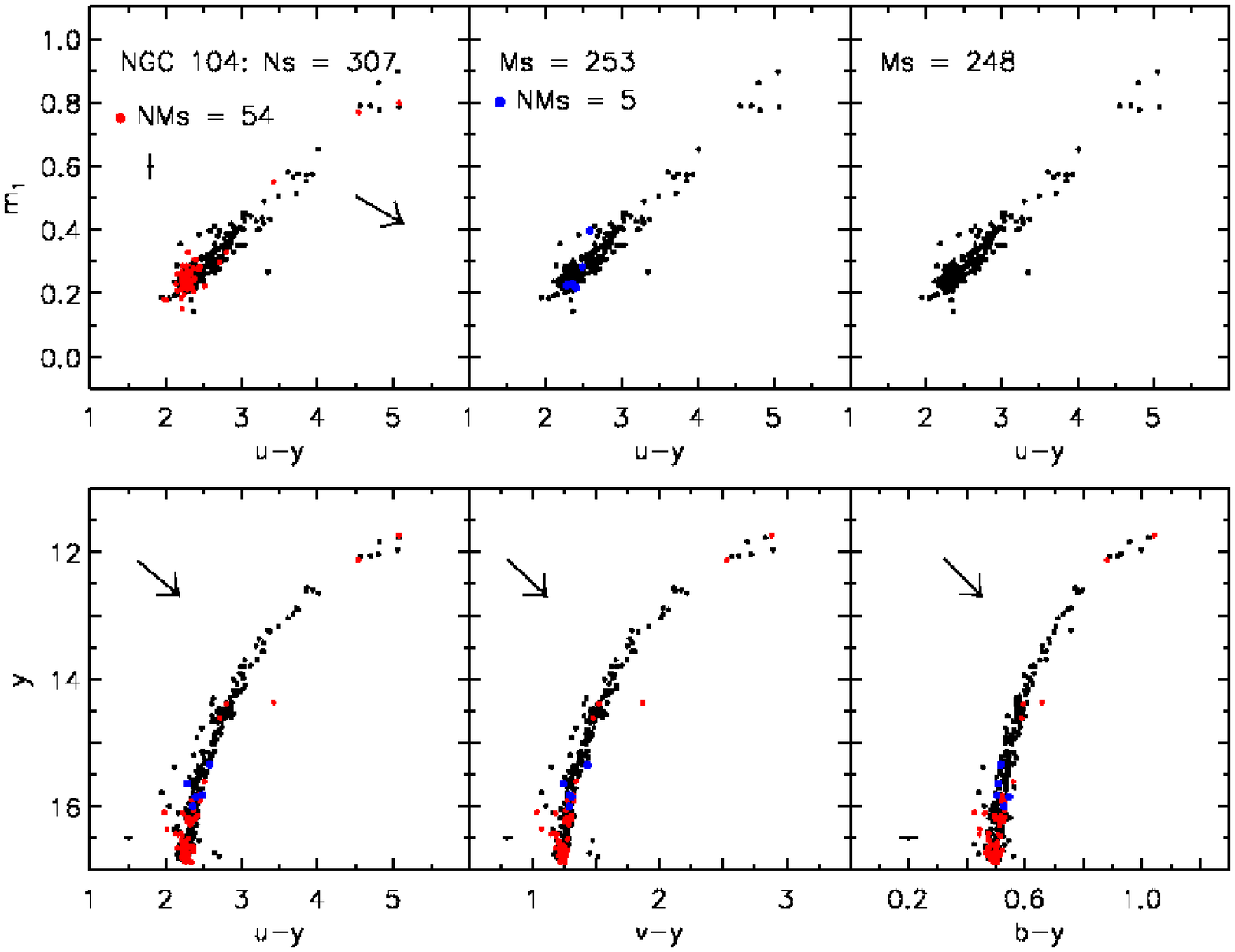}
\caption{Same as Fig.~3, but for RG stars in the GC NGC$\,$104. Only the 
RG stars with a proper motion velocity $\le 30~mas/yr$ were considered 
candidate cluster members.}
\end{center}
\end{figure}

Thus, we face the empirical evidence that the three metal-poor 
($\feh\le -1.65$) GCs---M92, NGC$\,$6397, M13---do not show clear 
evidence of candidate cluster RG stars with large $m_1$ values once 
the selection criteria were applied. On the other hand, the 
metal-intermediate ($\feh\sim -1.33$) GC NGC$\,$1851 shows evidence of 
a double sequence, while the more metal-rich ($\feh\sim -0.71$) GC 
NGC$\,$104 shows evidence of a larger spread in color after 
the selection criteria were applied. Obviously, this trend cannot be 
caused by photometric errors alone, since the uncertainties in the \strom 
magnitudes in the different clusters are smaller than 0.03~mag, and should 
not be very different from one cluster to the next.  Moreover, 
the error bars plotted in the top left panels display the uncertainty, 
summed in quadrature, for the faintest RG stars in each sample and they 
are smaller than the observed spread. A quantitative discussion of this 
trend with cluster metal abundance is beyond the aim of this investigation. 
In passing, we note that a large spread in carbon abundances has been 
spectroscopically measured among subgiant and main sequence (MS) stars of 
M13 (Briley, Cohen, \& Stetson 2002; Briley et al.\ 2004) and NGC$\,$104 
(Harbeck, Smith, \& Grebel 2003). A strong scatter in CN abundances 
has also been measured among RG (HR00), subgiant and MS stars of 
$\omega$ Centauri (Stanford et al.\ 2004; Kayser et al.\ 2006).   
However, we excluded the stars with large $m_1$ values from the sample 
of candidate cluster RG stars adopted to calibrate the \strom metallicity 
index. The reason is twofold: {\em i\/}) we are interested in a photometric 
proxy for the iron abundance;  {\em ii\/}) we still lack a firm explanation of 
the physical mechanisms that govern the occurrence of this spread in chemical 
composition. A detailed analysis of evolved and MS cluster stars showing a 
large spread in $m_1$ and in $c_1$ values will be addressed in a future 
paper.  

The empirical validations that we performed for the calibrating clusters 
by using the \umy\ colors have also been performed using the 
\vmy\ colors. We found the same results, so they will not 
be repeated here.     


\section{Empirical calibration of the \strom metallicity index}

\subsection{Multilinear regression fit}

In order to provide a metallicity calibration of the \strom index that 
can be applied to RG stars in different stellar environments we decided 
to derive new empirical metallicity-index--color (MIC) relations. 
In estimating the empirical MIC relations that correlate the iron abundance 
of RG stars to their metallicity index ($m_1$) and color index ({\it CI\/}), 
we dereddened the different samples. For each cluster 
we adopted the reddening value listed in Table~2 and then, according to 
Crawford \& Mandwewala (1976), we adopted $E(\bmy)= 0.74\, E(\bmv)$, 
$E(\vmy)=1.24\, E(\bmv)$, $E(\umy)= 1.79\, E(\bmv)$, and also 
$E(m_1)= -0.32\, E(\bmy)$. By using the reddening law from 
Cardelli et al.\ (1989) and $R_V=3.1$, we found $E(\bmy)= 0.70\, E(\bmv)$, 
$E(\vmy)= 1.33\, E(\bmv)$, $E(\umy)= 1.84\, E(\bmv)$, and 
$E(m_1)= -0.30\, E(\bmy)$. The two different sets agree quite-well 
within the typical 15\% uncertainty of current reddening laws 
(Rieke \& Lebofsky 1985).  
Together with the dereddened $m_1$ index ($m_{10}$, hereinafter $m_0$),
we also plan to derive independent empirical MIC relations for the 
reddening-free parameter $[m] = m_1\, +\, 0.32\,(\bmy)$. This \strom 
index was adopted to overcome deceptive uncertainties in clusters 
possibly affected by differential reddening. 

\begin{figure}[!ht]
\begin{center}
\label{fig11}
\includegraphics[height=0.6\textheight,width=0.75\textwidth]{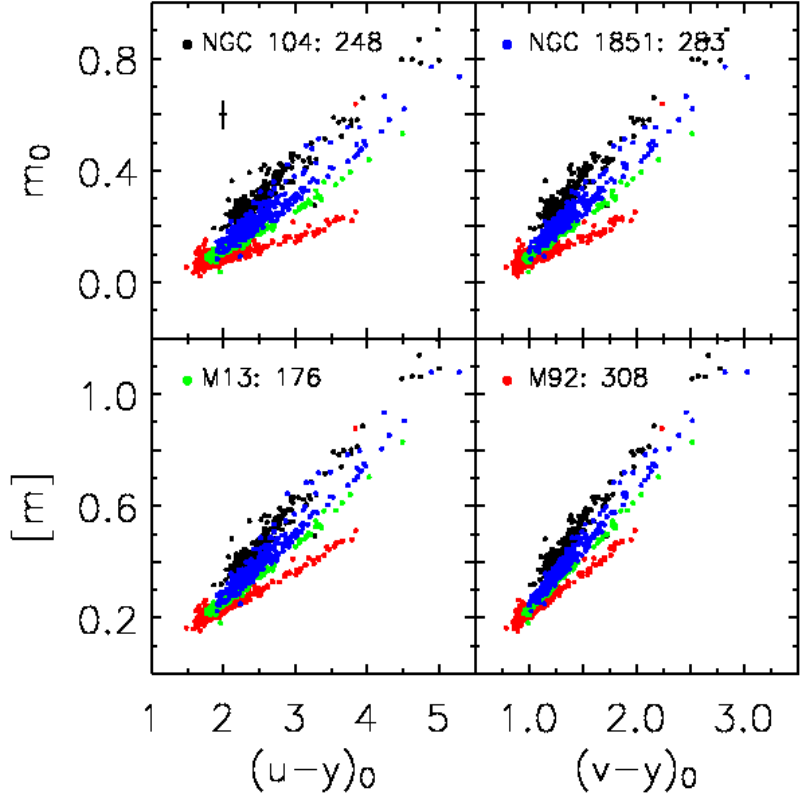}
\caption{Candidate RG stars for the four calibrating clusters plotted 
in different metallicity index-color planes. Dots of different colors mark 
RG stars of different clusters. The error bars in the top left panel account 
for uncertainties both in the photometry and in the reddening correction. 
The number of selected RG stars in each cluster are also labeled.}
\end{center}
\end{figure}

Fig.~11 shows the cleaned samples of candidate RGs for the four calibrating 
clusters (M92, M13, NGC$\,$1851, NGC$\,$104) in four different dereddened MIC 
planes. The error bars plotted in the top left panel display the photometric 
error budget for the dereddened color indices 
($\sigma(\umy)_0 \le 0.05$, $\sigma (\vmy)_0 \le 0.045$ mag) and for the 
metallicity indices ($\sigma (m_0) = \sigma ([m]) \le 0.05$ mag). 
Current errors are generous estimates, since they account for the 
uncertainties in reddening corrections and in the photometry. The 
latter errors were estimated in the faint magnitude limit of 
RG stars. Data plotted in Fig.~11 show two compelling pieces of empirical 
evidence: {\em i\/}) the $m_0/[m], CI_0$ relations are linear over a 
broad color range, namely $1.5\lesssim (\umy)_0 \lesssim 5.0$ and 
$0.85\lesssim (\vmy)_0\lesssim 3.0$; {\em ii\/}) the metallicity indices 
are well correlated with the cluster iron abundance, and indeed the 
four calibrating clusters present sharp and well-defined slopes. 
Note that observational findings (see Fig. 8 in ATT00) indicate that the
$m_1,\,(b-y)$ relation for cluster red giants is not linear over the entire 
color range ($0.4 \le (b-y) \le 1.1$).

For each cluster the RG sequence was uniformly sampled in steps of 0.15 mag 
in the color range $1.6<(\umy)_0<5.2$, and in steps of 0.1 mag in the color 
range $0.85<(\vmy)_0<3.05$. On the basis of these mean color bins, we 
also estimated for each cluster the corresponding mean $m_0$ and $[m]$ values 
and their errors. Hence, we applied a multilinear regression fit, by adopting 
the cluster metallicities listed in Table~2, to estimate the coefficients of 
the different MIC relations:

\begin{equation}
m_{0} = \alpha + \beta\,\feh + \gamma\, CI_0 +
\delta\,\feh\, CI_0
\end{equation}

where the symbols have their usual meaning. Similar relations have also been 
derived for the $[m]$ index. Note that the cluster metallicities adopted in 
the fit are on the Zinn \& West (1984) scale and based on the calcium triplet 
measurements provided by Rutledge et al.\ (1997). The coefficients of the fits, 
together with their uncertainties, for the four different MIC relations, are listed 
in Table~3. The multi-correlation parameters of the different relations 
are listed in column (6) and attain values very close to 1, thus supporting 
the use of the quadratic term involving metal abundance and color index. 
It is also noteworthy that the coefficients of the terms including the 
color index are systematically larger for the \vmy\ than for the \umy\  
color index. This means that {\em for fixed photometric errors\/}, the 
\umy\  color provides a better \feh\ determination than the \vmy\ color.  
However, accurate $u$-band measurements are generally much more difficult 
to obtain than $v$-band measurements. Therefore, {\em for fixed observing time\/}
the \vmy\ color may be more efficient than the \umy\  color. According to their 
definitions  $m_1 = (\vmb) - (\bmy)$ and $[m] = m_1 + 0.3 (\bmy)$, hence one can 
provide a metallicity determination by using only three measurements ($v,b,y$). 

\section{Validation of the new metallicity calibration}\label{validation}

In order to constrain the plausibility of the new empirical calibration
of the MIC relations we decided to apply them to five GCs for which
accurate and homogeneous \strom photometry, accurate absolute calibration,
and sizable samples of RG stars are available. They are NGC$\,$288, NGC$\,$362, 
NGC$\,$6752, NGC$\,$6397, and M71. The first three clusters are marginally affected 
by reddening and indeed $E(\bmv)\le 0.04$, while the last two have estimated reddening values of
$E(\bmv) = 0.18$ and $E(\bmv) = 0.31$, respectively. The chemical compositions
of these five clusters also cover more than 1 dex in iron abundance, namely 
from $\feh\sim -1.9$ (NGC$\,$6397) to $\feh\sim -0.7$ (M71, see Table~2).    

\begin{figure}[!ht]
\begin{center}
\label{fig12}
\includegraphics[height=0.75\textheight,width=0.8\textwidth,angle=90]{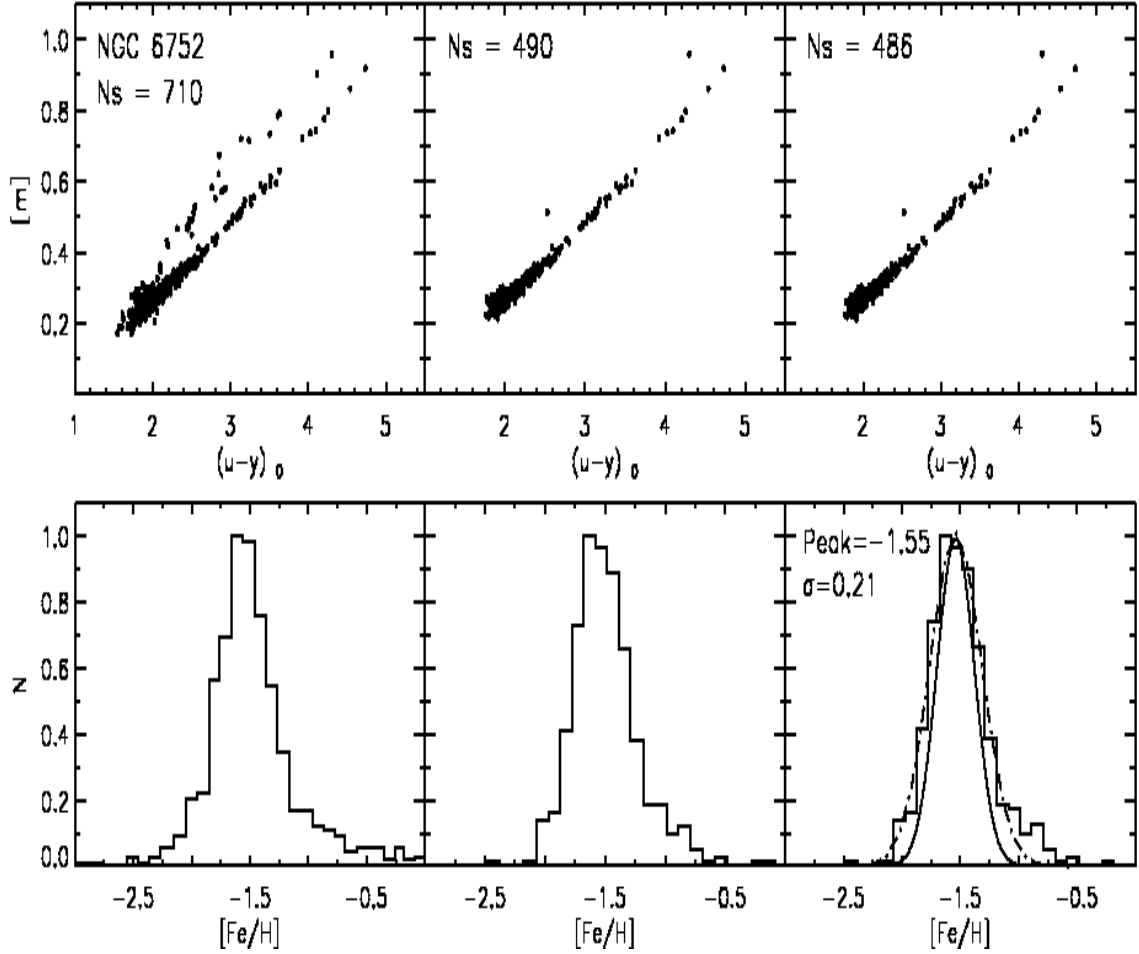}
\caption{Top: RG stars in NGC$\,$6752 plotted in the $[m], (\umy)_0$ plane (left). 
The middle panel shows in the same plane the candidate cluster RG stars after the 
selection in the $(\umj,\,\bmh)$ plane, while the right panel after the PM selection.
Candidate cluster RG stars were selected according to optical-NIR colors 
($\Delta (\umj) \le 1.5\, \sigma_{\umj}$) and proper motion velocity ($\le 25\, mas/yr$).
Bottom: metallicity distribution obtained using the new $[m], (\umy)_0$ MIC relation  
(see Table~3) and applied to the RG stars of the top panels. 
The distribution of the cleaned sample (right bottom panel) was fitted with a 
Gaussian (dashed-dotted line) with a peak value of $\feh_{phot} = -1.55$ dex, 
and a dispersion $\sigma = 0.21$ dex. The solid line displays the Gaussian 
accounting for photometric errors and uncertainties in reddening corrections.}
\end{center}
\end{figure}

In order to estimate the mean metallicities of the GCs adopted to validate the 
new MIC relations we adopted the same approach devised for the calibrating GCs. 
We selected {\em bona fide\/} RG stars by using the optical-NIR color-color plane 
and when available the proper-motion data (NGC$\,$288, NGC$\,$6752, NGC$\,$6397).   
The top panels of Fig.~12 show RG stars in NGC$\,$6752 plotted in the
($[m], (\umy)_0$) plane: in the left panel is plotted the entire sample of RG 
stars, while the middle one shows the candidate cluster RGs after the optical-NIR 
color--color selection was applied, and the right one shows the sample remaining after the PM selection.
The bottom panels of Fig.~12 show the metallicity distributions obtained for 
the three different RG samples using the new MIC relation between $[m]$ and $(\umy)_0$ 
(see Table~3). The  distribution of the cleaned sample was fitted with a Gaussian function, 
with a peak value of $\feh_{phot} = -1.55$ dex and a dispersion of 
$\sigma = 0.21$ dex (dashed-dotted line). This cluster metallicity estimate 
agrees quite-well with the spectroscopic result 
($\feh_{spec} = -1.54\pm0.09$). The same conclusion applies to the metallicity 
estimates based on the other MIC relations (see Table~4), and indeed the cluster 
metallicity estimates range from $\feh_{phot} = -1.56\pm0.19$ ($m_0, (\umy)_0$) 
to $\feh_{phot} = -1.66\pm0.18$ ($m_0, (\vmy)_0$).     
The dispersion of the current metallicity estimates is not intrinsic but mainly 
caused by photometric errors ($\sigma = 0.15$ dex), uncertainties in the 
coefficients of the MIC relations ($\sigma = 0.02$ dex) and in the reddening 
correction ($\sigma = 0.04$ dex).
The solid line plotted in the bottom right panel of Fig.~12 shows a Gaussian function
that accounts for the entire error budget. The two dispersions attain quite similar 
values over the entire range, thus supporting the evidence that the spread in 
metal abundances is due only to intrinsic errors.  

It should be remembered, however, that the lines of constant metal abundance 
fan apart (Fig.~11), being much more
widely separated for the most luminous, coolest giants in the various clusters,
and much closer together for stars near the base of the clusters' giant branches.
Thus, for constant photometric errors and uncertainties in the reddening corrections,
luminous giants will provide {\it much\/} more precise metallicity estimates than
faint ones.  Furthermore, uncertainties in the coefficients of the MIC relations
will affect intermediate-luminosity giants least, and the brightest and faintest
giants more.  Finally, the faintest giants are, as it is evident from Fig.~12, 
subject to perceptibly larger photometric uncertainties than brighter ones, further
reducing their value for metallicity estimates in the present case.  The solid
curve in the bottom-right panel of Fig.~12 should not properly be conceived, then,
as a single Gaussian function, but rather as a superposition of a few narrow
functions for the few brightest giants, slightly broader functions for the
more numerous intermediate-luminosity giants, and many much broader functions
for the faintest giants.  The curve and stated dispersion of $\sigma(\feh) = 0.21\,$dex
should therefore be understood as {\it representative\/} rather than as
{\it mathematically correct\/}.

Interestingly enough, we find a similar remarkable agreement also for the other 
four GCs we adopted to validate current MIC relations. The comparison between the 
mean photometric metallicity estimates listed in Table~4 and the spectroscopic 
measurements given in Table~2 indicates that they agree with each other within 
$1\sigma$ errors. It is noteworthy that the accuracy of photometric metallicity 
estimates applies not only to metal-intermediate clusters with low reddening 
corrections (NGC$\,$288), but also to more metal-rich clusters with high reddening 
(M71). The metallicity estimates for the GC M71 show, 
as expected, a larger spread but this is still within the errors and is due primarily to the 
larger uncertainty in the reddening correction. Note that an uncertainty of 
$\sim$ 0.05 mag in $E(\bmv)$ causes a systematic uncertainty in \feh\ of $\sim$ 0.1 dex 
and of $\sim$ 0.08 dex using the MIC relations including the \umy\  or the \vmy\  
color, respectively. 

\begin{figure}[!ht]
\begin{center}
\label{fig13}
\includegraphics[height=0.6\textheight,width=0.5\textwidth]{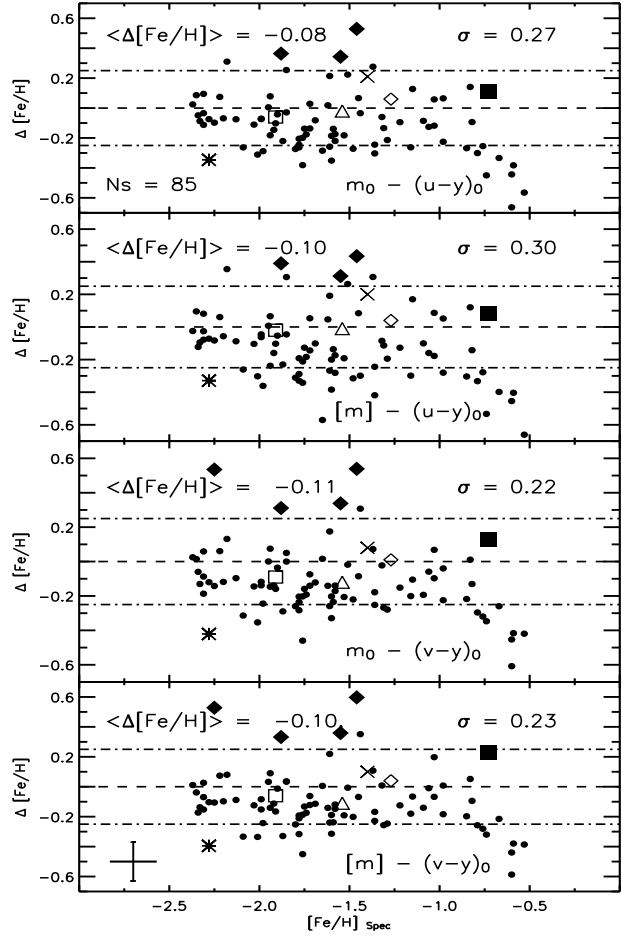}
\caption{Difference between photometric and spectroscopic metallicities,
$\Delta \feh = (\feh_{\hbox{\footnotesize phot}} - \feh_{\hbox{\footnotesize spec}})$, plotted versus
$\feh_{\hbox{\footnotesize spec}}$ for the 85 field stars of the ATT94 and ATT98 sample. 
The four panels display the four empirical MIC relations whose coefficients 
are listed in Table~3.  The filled diamonds mark the CH-strong stars, namely 
HD $\,$55496, HD$\,$135148, BD-012582, BD+042466, and CD-621346, while the 
asterisk marks the peculiar star HD$\,$84903. Different symbols 
show the photometric metallicity estimates of the five validation GCs: 
-- empty square: NGC$\,$6397, -- empty triangle: NGC$\,$6752, -- cross: NGC$\,$288, 
-- empty diamond: NGC$\,$362, -- filled square: M71. The error bars in the bottom panel 
display the mean error for the spectroscopic abundance measurements.}
\end{center}
\end{figure}

To further constrain the new calibration of the MIC relations, we used 
them to estimate the metallicity of NGC$\,$6397 by adopting the \strom photometry 
of ATT00 for this cluster. This photometry is on the \strom system of Bond (1980) 
and we adopted the color equations given by Olsen (1995, hereinafter O95)
to transform the ATT00 photometry into the  \strom system of Olsen (1993, 
hereinafter O93). By applying the calibration to RG stars in common between 
their photometry and ours, we obtained two very similar metallicity 
distributions. 
The estimated mean iron abundance is $\feh_{\hbox{\footnotesize ours}} = -1.99$ dex 
with an intrinsic dispersion of $\sigma_{\hbox{\footnotesize ours}}=0.21$ dex 
according to the ($[m], (\umy)_0$) MIC relation and 
$\feh_{\hbox{\footnotesize ours}} = -2.02$ with 
$\sigma_{\hbox{\footnotesize ours}}= 0.19$ dex according to the 
($[m], (\vmy)_0$) MIC relation.   
By using the same MIC relations and the ATT00 photometry we find: 
$\feh_{\hbox{\footnotesize ATT00}} = -2.24$ dex 
with $\sigma_{\hbox{\footnotesize ATT00}} = 0.17$ dex and 
$\feh_{\hbox{\footnotesize ATT00}} = -2.22$ dex
with $\sigma_{\hbox{\footnotesize ATT00}} = 0.18$ dex, respectively.
The above differences are due to a difference in the photometric calibration. We found
that the relative difference in the $\bmy$ color ((our-ATT00) vs our) presents a
linear trend, that is of the order of -0.04 mag for $\bmy\sim 0.5$ and of the order
of 0.02 mag for $\bmy\sim 1.0$. A zero point difference of 0.01 mag was found for
the $m_1$ index. On the other hand, we found that the $c_1$ index presents a parabolic
trend that is of the order of -0.01 mag for $c_1\sim 0.3$ and of the order of -0.07 mag
for $c_1\sim0.6$. The absolute calibration of \strom photometry in this GC deserves
further investigations.

Together with the validation based on GCs we decided to test the accuracy 
of the new empirical MIC relations using a sample of field RG stars. 
Table~4 of Anthony-Twarog \& Twarog (1998, hereinafter ATT98) gives 121 
field stars with published $uvby$ photometry (SN88, O93, ATT94, ATT98) and 
spectroscopic measurements. This sample
we complemented with 15 stars from ATT94 for which are available $uvby$
photometry and spectroscopic measurements. The $u$-band photometry for
the star BD+032782 was retrieved from the catalog by Hauck \& Mermilliod (1998).
We ended up with a sample of 137 field stars. The metallicity range covered by current
MIC relations is $-2.2<\feh<-0.7$, but we select the stars with $-2.4<\feh<-0.5$ to
account for current uncertainties in spectroscopic abundances and in the GC metallicity
scale (Kraft \& Ivans 2003). We rejected 24 stars, because they are either less metal-poor
(19 stars with $\feh<-2.4$) or more metal-rich (5 stars with $\feh>-0.5$) than the quoted
metallicity range. We excluded the star HD$\,$251611 because it is bluer (according to 
Hauck \& Mermilliod 1998, the apparent color for this object is $\bmy = 0.452$, and the 
reddening $E(b-y)=0.035$, transformed into the O93 system, its unreddened color is 
$(\bmy)_0 = 0.388$)
than the color range covered by current MIC relations ($0.425 \le (\bmy)_0 \le 1.05$).
We also rejected other 25 stars, because 22 of them are Red Horizontal Branch stars,
two are dwarf stars (BD+133683 and CD-310622, $\log g = 4$) and one is a sub-giant
star (HD$\,$18907). Two stars (BD+122546, GPEC2643) were excluded because the $u$-band
photometry was not available in the literature. As a whole, we ended up with a sample
of 85 stars. Table~6 lists the selected stars together with the spectroscopic measurements
collected by ATT94 and ATT98. The iron abundances of this sample were transformed by ATT98
into the the metallicity scale of Kraft et al.\ (1992), that is consistent with the
Zinn \& West metallicity scale. The $y$ magnitudes and the \bmy\ colors provided by ATT98
are on the photometric system of O93, while the $m_1$ indices have been transformed into
the \strom system of Bond (1980). By adopting the color equations given by O95, we
transformed the ATT98 photometry into the system of O93. The reddening estimates for these
stars have been estimated by ATT98 (see also ATT94 for more details).

We plotted the difference between the photometric and the spectroscopic
metallicity for the 85 field RG stars as function of their spectroscopic
iron abundances in Fig.~13. The difference between photometric estimates and
spectroscopic measurements for the 5 GCs adopted to validate the calibration
is also shown. Data plotted in this figure show that on average there is a
systematic shift of $\sim$ 0.1 dex towards more metal-poor values for the four
different MIC relations. In spite of this systematic difference, the intrinsic
dispersion is smaller than 0.3 dex and it is mainly due to photometric, reddening,
and spectroscopic errors. The error bars in the bottom panel of Fig.~13 displays
the mean error for the spectroscopic abundance measurements
($\sigma(\feh_{\hbox{\footnotesize spec}}) \sim$ 0.13), estimated as the average of
both the internal dispersion about the mean $\feh$ measurements and the uncertainty
due to the transformations to the standard metallicity scale (see column 8 in Table~2
and column 7 in Table~4 of ATT98). If we remove the five CH-strong stars (HD $\,$55496,
HD$\,$135148, BD-012582, BD+042466, CD-621346, marked with diamonds) and the peculiar star
HD$\,$84903 (marked with an asterisk), studied by Smith, Dupree, \& Churchill (1992), and
showing chromospheric emission in the Ca II K line core (see ATT98), the dispersion of the
residuals is  $\sigma \lesssim$ 0.2 dex (see Table~5).

The photometric metallicities for four metal-rich stars (CP-570680, BD-182065, HD 37160
and HD 6833) with ($\feh >$ -0.70) are systematically more metal-poor by 0.3-0.4 dex than
spectroscopic measurements. H00 also estimated for CP-570680 and BD-182065 a photometric
abundance $\sim$ 0.2-0.3 dex more metal-poor than the spectroscopic values. This discrepancy
might be due to the fact that they are slightly more metal-rich than the metallicity range
covered by current MIC relations. They deserve a more detailed spectroscopic investigation
to assess whether they possess distinct peculiarities.

A detailed discussion concerning the metallicity distribution obtained using 
the new MIC relations and the comparison with similar MIC relations available 
in the literature is presented in \S 7.  

\section{Semi-empirical and theoretical framework}\label{theo}

In order to constrain the occurrence of possible systematic errors in the new 
empirical MIC relations, we decided to derive independent MIC relations using the 
homogeneous and updated set of evolutionary models recently constructed by 
Pietrinferni et al.\ (2004) for a scaled-solar heavy element mixture and 
by Pietrinferni et al.\ (2006) for an $\alpha$-enhanced ($\afe=0.4$)
mixture. This extended set of stellar isochrones and 
zero-age horizontal-branch (ZAHB) models offers a twofold advantage: 
{\em i\/}) they cover a broad range in metal abundance; 
{\em ii\/}) the evolutionary models have been constructed by adopting the 
same input physics for both scaled-solar and $\alpha$-enhanced chemical 
compositions. Theoretical predictions have been transformed into the 
observational plane by adopting a new set of bolometric corrections (BCs) 
and CTRs discussed in detail by Pietrinferni et al.\ (2004, 2006) and 
by CK06\footnote{The complete 
set of BCs and CTRs is available at the URL 
http://wwwuser.oat.ts.astro.it/castelli/colors.html} 
that properly account for the difference in the adopted heavy element 
mixtures.

\begin{figure}[!ht]
\begin{center}
\label{fig14}
\includegraphics[height=0.55\textheight,width=0.65\textwidth]{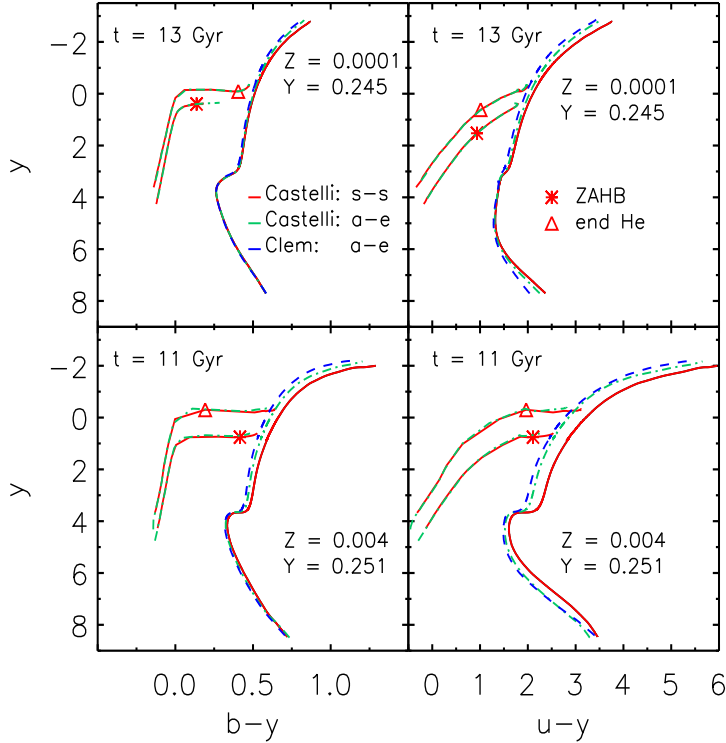}
\caption{Comparison between scaled-solar (solid red lines) and 
$\alpha$-enhanced (dashed-dotted green lines) isochrones for 
two different chemical compositions and stellar ages (see labeled 
values). Evolutionary predictions have been transformed into the 
observational plane by adopting the CTRs provided by CK06. The dashed 
blue lines display the same $\alpha$-enhanced isochrones, but 
transformed using the semi-empirical CTRs provided by CVGB04. 
Left and right panels show the comparison in the ($y$,\,\bmy) and 
($y$,\,\umy) CMDs. The ZAHBs (asterisks) and the exhaustion of central 
He-burning (triangles) for the two different chemical compositions 
are also shown.}
\end{center}
\end{figure}

Fig.~14 shows the comparison between scaled-solar (solid red lines) 
and $\alpha$-enhanced (dashed-dotted green lines) isochrones for two 
different chemical compositions and stellar ages, namely 
Z=0.004, Y=0.251, t = 11 Gyr (bottom panels) and Z=0.0001, Y=0.245, 
t = 13 Gyr (top panels) in two different CMDs ($y$,\,\bmy, left panels) 
and ($y$,\,\umy, right panels). These models
were transformed into the \strom observational plane by adopting 
the CTRs provided by CK06 for the same heavy element mixture used in 
the evolutionary computations. In order to constrain the occurrence  
of deceptive uncertainties affecting the fully theoretical CTRs -- mainly 
in the low-$T_{\hbox{\footnotesize eff}}$ regime -- we decided to 
transform the $\alpha$-enhanced stellar isochrones by also adopting 
the semi-empirical CTRs recently provided by CVGB04. This set of 
BCs and CTRs is based on atmosphere models that account for an 
$\alpha$-enhanced mixture at lower effective temperatures 
($T_{\hbox{\footnotesize eff}} \le 8000\,$K). However, the 
predicted colors were empirically calibrated using a large sample of 
field stars. The interested reader is referred to the aforementioned paper 
for more details concerning the approach adopted by CVGB04 to 
calibrate and validate their CTRs. The dashed blue lines plotted 
in Fig.~14 represent the $\alpha$-enhanced isochrones transformed 
using these semi-empirical CTRs. 

Data plotted in the top panels of Fig.~14 show that the difference 
between scaled-solar and $\alpha$-enhanced isochrones transformed by 
adopting both theoretical and semi-empirical CTRs is marginal in the 
metal-poor regime. On the other hand, in the more metal-rich regime, 
the $\alpha$-enhanced isochrones attain bluer colors than scaled-solar 
ones. The difference in ($y$,\,\bmy) CMD becomes evident only for  
effective temperatures cooler than the turnoff region, while in 
the ($y$,\,\umy) CMD the $\alpha$-enhanced isochrones are systematically 
bluer than scaled-solar ones. On the other hand, the ZAHBs (asterisks) 
and the exhaustion of core-helium burning (triangles) constructed by 
adopting $\alpha$-enhanced (dashed-dotted green line) and scaled-solar 
(solid red line) compositions do show a negligible difference. These 
findings support the early results by Cassisi et al.\ (2004) and 
by Pietrinferni et al.\ (2006) concerning the dependence of 
intermediate-band CTRs on the abundance of heavy elements. 

\begin{figure}[!ht]
\begin{center}
\label{fig15}
\includegraphics[height=0.80\textheight,width=0.95\textwidth]{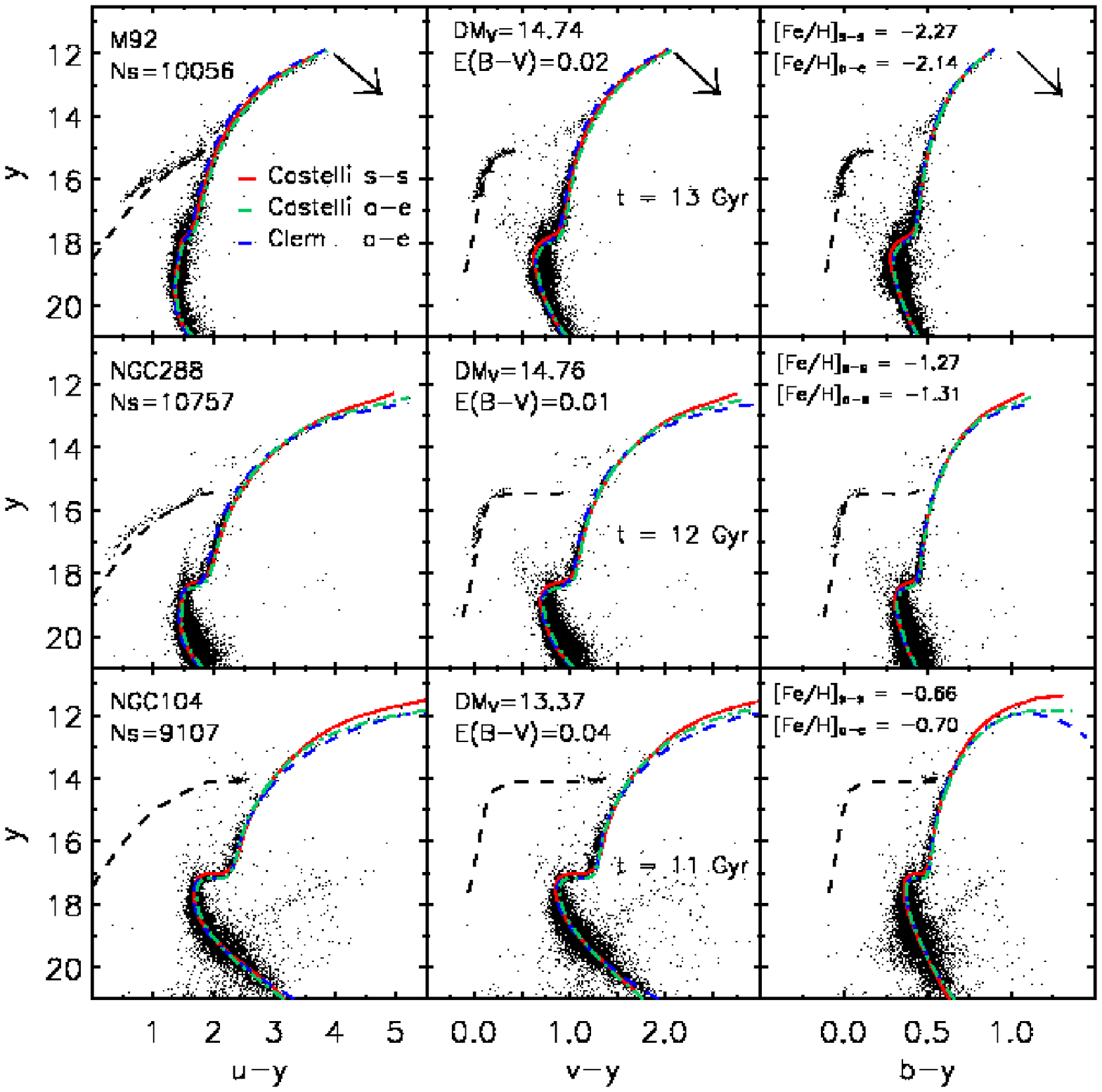}
\caption{\strom CMDs for three globular clusters with 
different metal abundances, namely M92 (top), NGC$\,$288 (middle), 
and NGC$\,$104 (bottom). From left to right the panels display the CMDs 
in ($y$,\,\umy), ($y$,\,\vmy), and ($y$,\,\bmy).
The red solid and the dashed-dotted green lines show the scaled-solar 
($\feh_{s-s}$) and the $\alpha$-enhanced ($\feh_{a-e}$) isochrones 
for the labeled metallicities and cluster ages. These isochrones were 
transformed into the observational plane by using the theoretical CTRs 
provided by CK06. The dashed blue lines display the same $\alpha$-enhanced
isochrones, but transformed using the semi-empirical CTRs provided by CVGB04.   
The dashed black lines display the scaled-solar ZAHBs. The reddening 
vectors for the three colors are shown in the top panels.}
\end{center}
\end{figure}

In order to validate the current transformations we selected three globular 
clusters that cover a broad metallicity range, namely M92, NGC$\,$288, 
and NGC$\,$104. Fig.~15 shows the comparison between theory and 
observations in three different CMDs: ($y$,\,\umy), ($y$,\,\vmy), and ($y$,\,\bmy). 
The fit between the different sets of stellar isochrones and the 
observations was performed using the same distance moduli:  
$DM_0 = 14.68\pm0.07$, which for $E(\bmv)=0.02$ becomes 
$DM_V = 14.74\pm0.07$ for M92 (Carretta et al.\ 2000); 
$DM_0 = 14.73\pm0.10$, which for $E(\bmv)=0.01$ becomes 
$DM_V = 14.76\pm0.10$ for NGC$\,$288 (Ferraro et al.\ 1999), and
$DM_0 = 13.25\pm0.06$, which for $E(\bmv)=0.04$ becomes 
$DM_V = 13.37\pm0.06$ for NGC$\,$104 (Percival et al.\ 2002). 
The reddening corrections for the different \strom colors were 
estimated using the relation given in \S 4.1. The iron abundance 
of both scaled-solar ($\feh_{\hbox{\footnotesize s-s}}$) and $\alpha$-enhanced 
($\feh_{\hbox{\footnotesize a-e}}$) isochrones are also labeled. 

Data plotted in the top and in the middle panels of Fig.~15 reveal that the 
difference between scaled-solar and $\alpha$-enhanced isochrones is negligible 
in the metal-poor and in the metal-intermediate regime in these 
\strom CMDs. In the more metal-rich regime (bottom panels), the scaled-solar 
isochrones in the ($y$,\,\umy) and ($y$,\,\vmy) CMDs, transformed using the 
theoretical CTRs of CK06, appear slightly redder than observed when moving 
toward cooler RG stars. On the other hand, the two different sets of 
$\alpha$-enhanced isochrones present a very similar trend. In order to 
account for possible subtle uncertainties in the CTRs, the stellar 
isochrones being the same, we decided to adopt in the following the 
scaled-solar isochrones transformed using the theoretical CTRs by CK06 
and the $\alpha$-enhanced isochrones transformed with the semi-empirical 
CTRs by CVGB04. 

\begin{figure}[!ht]
\begin{center}
\label{fig16}
\includegraphics[height=0.75\textheight,width=0.75\textwidth]{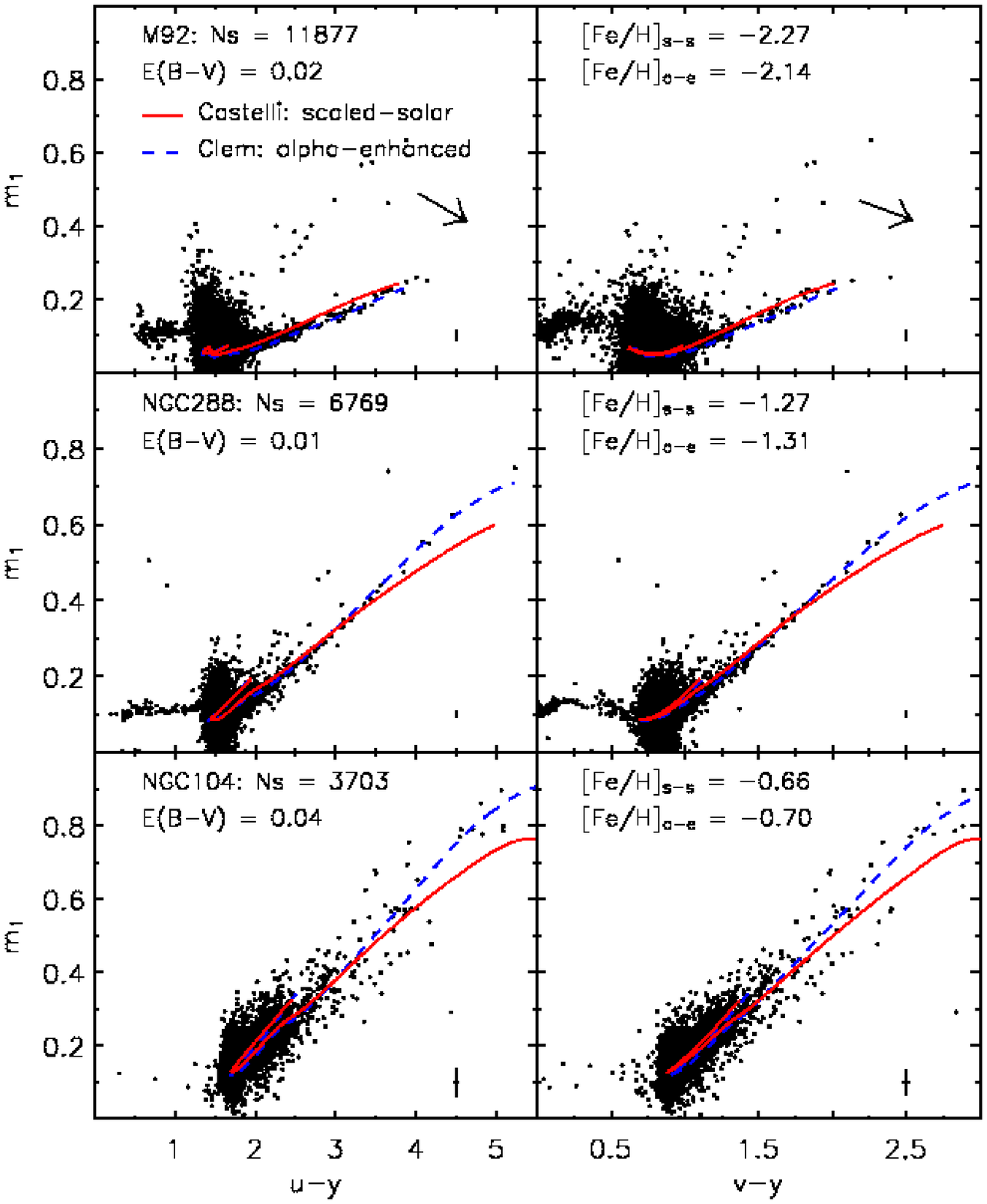}
\caption{Stars of the same globular clusters of Fig.~15 (M92, NGC$\,$288, 
NGC$\,$104), but plotted in the ($m_1$,\,\umy) (left panels) and 
($m_1$,\,\vmy) (right panels) color-color planes. Solid red lines display 
scaled-solar isochrones at fixed cluster age --$t = 12$ Gyr-- and for 
the labeled metallicities ($\feh_{s-s}$). The blue dashed lines display 
the $\alpha$-enhanced isochrones for the same cluster age the labeled 
metallicities ($\feh_{a-e}$), but they have been transformed using 
the CTRs by CVGB04. The reddening vector for the two colors is shown 
in the top panels. The error bars plotted in the bottom panels account 
for the photometric errors at the base of the cluster RGB.}
\end{center}
\end{figure}

As a further test of the intrinsic accuracy of the adopted CTRs, Fig.~16 
shows the comparison between selected isochrones and the same globular 
clusters as Fig.~15, but in the ($m_1$,\,\umy) and ($m_1$,\,\vmy) color-color 
planes. Data plotted in this figure display a reasonable agreement 
over the metallicity range covered by the selected clusters in the two 
different color-color planes.

\subsection{Semi-empirical and theoretical metallicity calibrations}

We performed two different metallicity calibrations adopting both 
the scaled-solar isochrones transformed into the \strom colors
using the CTRs by CK06 and the $\alpha$-enhanced ones transformed 
with the CTRs by CVGB04. For the scaled-solar and the $\alpha$-enhanced 
mixtures, we adopted eight metallicities, namely Z=0.0001, 0.0003, 0.0006, 
0.001, 0.002, 0.004, 0.008, 0.01.
These Z values correspond to the global abundance of heavy elements in the 
chemical composition mixture. This means that for a scaled-solar mixture, 
for which  $\feh=\mh$, the value of the iron content was obtained by 
using the relation $\feh=\log{(\zx)}-\log{(\zx)_\odot}$, where X is 
the abundance by mass of hydrogen and ${(\zx)_\odot}=0.0245$ is the 
solar iron abundance. For the $\alpha$-enhanced mixture 
this relation holds for the global metallicity \mh. Therefore, the corresponding
iron content was obtained by using the relationship given by 
Salaris et al.\ (1993): $\mh = \feh + \log(0.638f + 0.362)$
where $\log(f) = \afe$, and $\afe = 0.4$ is the 
$\alpha$-enhancement (Gratton et al.\ 2004).

To select the $m_1$ and $[m]$ values along the individual isochrones 
we followed the same approach adopted for the empirical calibration.
In particular, we selected 25 points in the color range 
$1.6<\umy<5.2$, with a step in color of 0.15 mag, and 23 points 
in the color range $0.85<\vmy<3.05$, with a step in color of 
0.1 mag. Then a multilinear regression fit was performed to 
estimate the coefficients (see Table~3) of the new MIC relations 
for the $m_1$ and the $[m]$ indices.

As a first step to validate the new theoretical and semi-empirical 
MIC relations we estimated the mean iron abundances of the nine  
globular clusters we adopted to calibrate and validate the empirical 
MIC relations (see Table~4). Fig.~17 shows that the photometric abundances 
based on the theoretical MIC relations are on average $0.11\pm0.03$ dex 
more metal-poor than spectroscopic measurements. Although, these 
estimates are still within the $1\sigma$ uncertainties of theoretical 
and empirical estimates, the systematic difference is clear and becomes 
of the order of 0.2 dex for a few metal-intermediate clusters.  

\begin{figure}[!h]
\begin{center}
\label{fig17}
\includegraphics[height=0.6\textheight,width=0.50\textwidth]{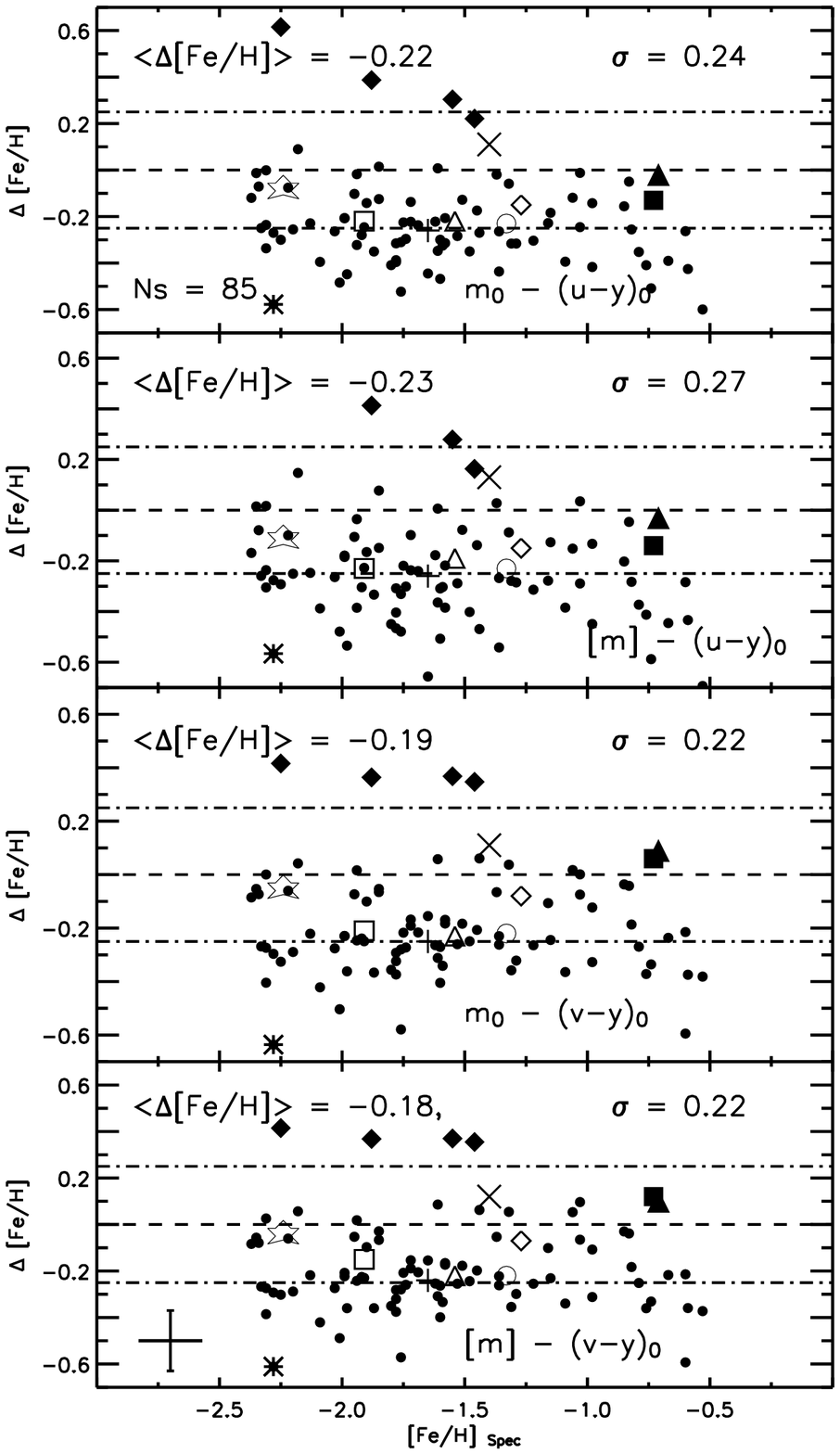}
\caption{Same as Fig.~13, but the photometric iron abundances are based
on the theoretical MIC relations listed in Table~3. The different symbols 
display the photometric metallicity estimates of nine globular clusters:  
-- star: M92, -- open square: NGC$\,$6397, -- plus: M13; 
-- open triangle: NGC$\,$6752, -- cross: NGC$\,$288, 
-- open circle: NGC$\,$1851, -- open diamond: NGC$\,$362, 
-- filled square: M71, -- filled triangle: NGC$\,$104.}
\end{center}
\end{figure}

In order to constrain this discrepancy we also applied the theoretical MIC relations
to estimate the iron abundances of the 85 field stars of the ATT94 
and ATT98 sample. Fig.~17 shows
the difference between the photometric and the spectroscopic iron abundances versus the
spectroscopic measurements. We find, as for the empirical calibration, that on average
the iron abundances appear to be shifted towards more metal-poor values,
$<\Delta(\feh_{\hbox{\footnotesize theo}} -
\feh_{\hbox{\footnotesize spec}})>\, \simeq -0.20\pm0.02$ dex,
even though the dispersion of the residuals is $\sigma \lesssim$ 0.27 dex.
Once we remove the five CH-strong stars and the peculiar star HD$\,$84903 the
dispersion of the residuals is $\sigma \lesssim$ 0.2 dex (see Table~5).

On the other hand, the photometric iron abundances for the nine 
globular clusters based on the semi-empirical MIC relations 
(see Table~4) agree quite well with the spectroscopic measurements. 
The mean difference is $0.04\pm 0.03$ dex, i.e., within current empirical 
and theoretical uncertainties. 

\begin{figure}[!ht]
\begin{center}
\label{fig18}
\includegraphics[height=0.6\textheight,width=0.5\textwidth]{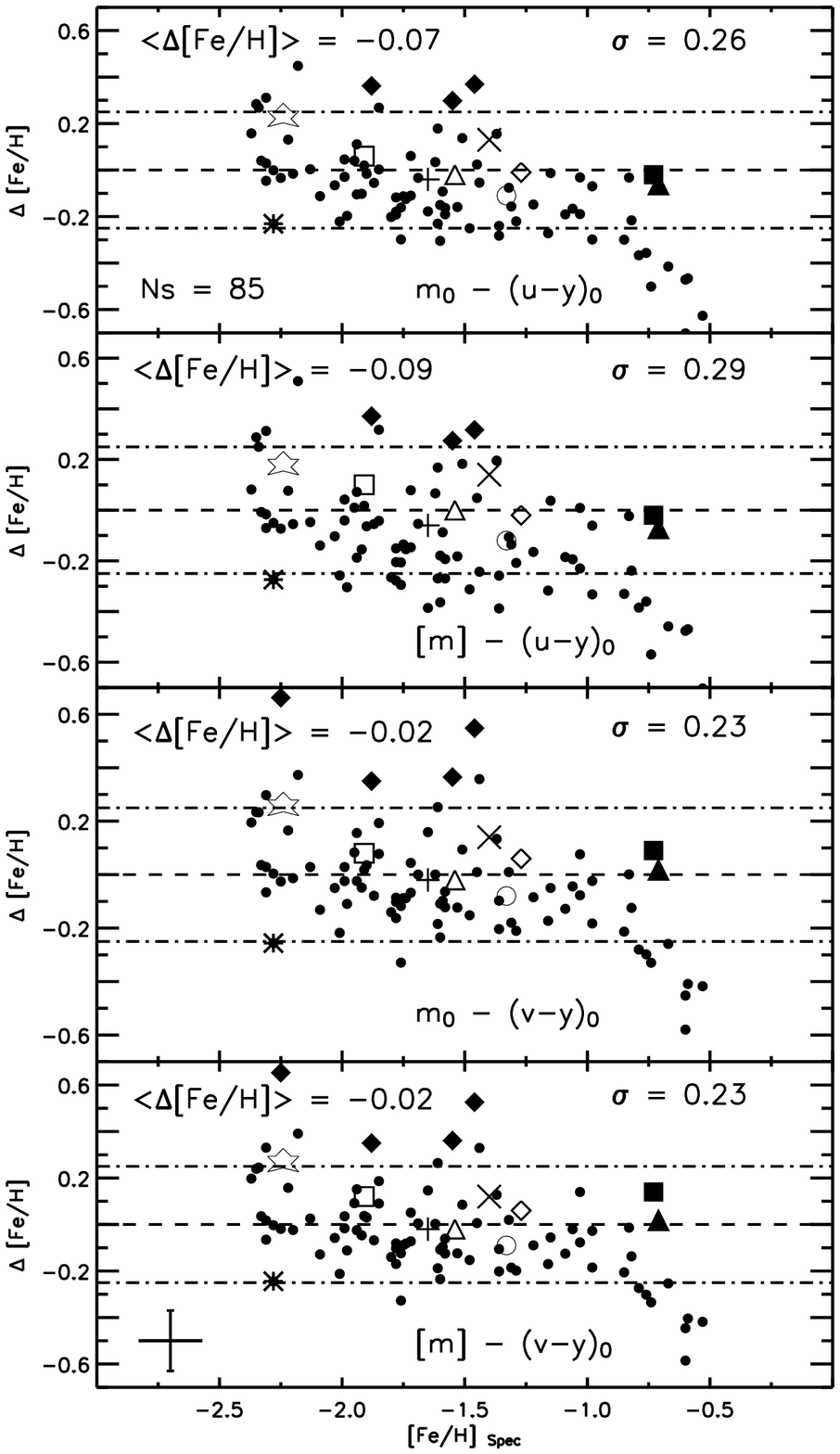}
\caption{Same as Fig.~17, but the photometric iron abundances are based
on the semi-empirical MIC relations listed in Table~3.}
\end{center}
\end{figure}

The differences between photometric abundances estimated with the semi-empirical
MIC relations and the spectroscopic measurements for the nine clusters are
plotted in Fig.~18, together with the 85 field stars of the ATT94 and ATT98 sample.
Data plotted in this figure show that photometric and spectroscopic abundances
agree quite well, and indeed no clear trend with iron abundance is present.
Moreover, the  difference on average is  $-0.05\pm 0.03$ dex with $\sigma \lesssim 0.3$ dex.
Once we remove the five CH-strong stars and the peculiar star HD$\,$84903 the
dispersion of the residuals is $\sigma \lesssim$ 0.2  dex. The mean of the residuals,
$<\Delta(\feh_{\hbox{\footnotesize phot}} - \feh_{\hbox{\footnotesize spec}})>$,
and the relative dispersion given by the different MIC calibrations are listed
in Table~5.

\begin{figure}[!ht]
\begin{center}
\label{fig19}
\includegraphics[height=0.6\textheight,width=0.5\textwidth]{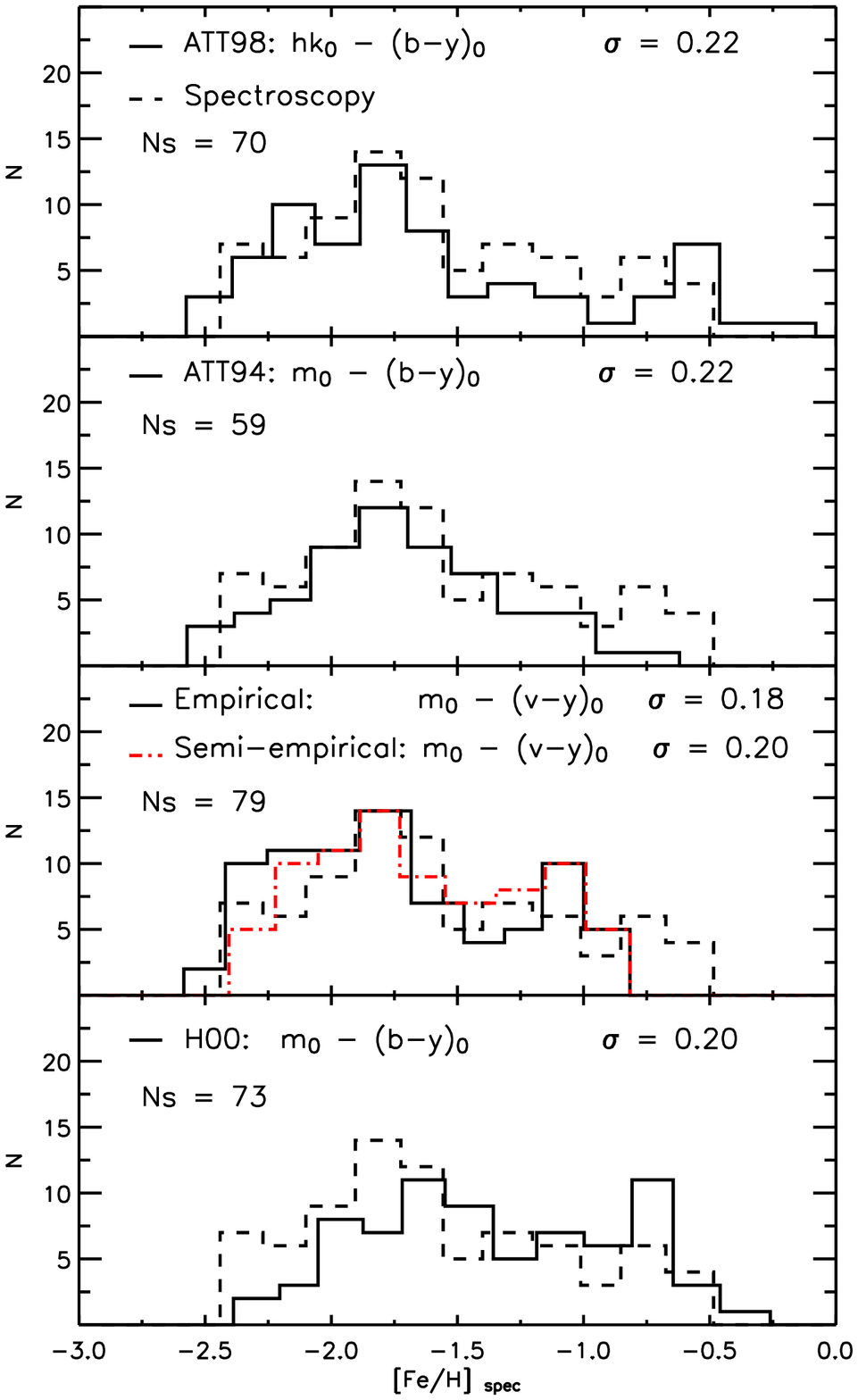}
\caption{Comparison between the spectroscopic metallicity distribution 
based on the 79 field stars from the ATT94 and ATT98 sample (dashed line)
and the photometric metallicity distributions based, from top to bottom, 
on the ATT98 empirical MIC relation --$hk_0, (\bmy)_0$-- (solid line, top panel); 
the ATT94 empirical MIC relation --$m_0, (\bmy)_0$-- (solid line, second panel);
our empirical --$m_0, (\umy)_0$-- (solid line, third panel)
and semi-empirical ($m_0,(\umy)_0$)-- (dashed-dotted red line, third panel) 
MIC relation;  the H00 empirical MIC relation --$m_0, (\bmy)_0$-- (solid line, 
bottom panel).}
\end{center}
\end{figure}

As a final test we decided to compare the spectroscopic metallicity
distribution for 79 stars, the five CH-strong stars and the peculiar star HD$\,$84903
were not included, of the ATT94 and ATT98 sample with the photometric metallicity 
distributions based on current empirical, theoretical and semi-empirical MIC relations, 
and on
similar MIC relations available in the literature. In order to detect the occurrence
of possible systematic errors affecting current MIC relations, we selected
several MIC relations based either on different colors or on
different metallicity indices. The top panel of Fig.~19 shows the comparison between
the spectroscopic metallicity distribution (dashed line) and the metallicity distribution
obtained with the empirical MIC relation based on the $hk$ index ($hk_0, (\bmy)_0$)
provided by ATT98 for 70 out of the 79 RG stars (see column 3 in Table~6). Nine stars 
have been excluded because
they do not have $hk$ photometry (HD$\,$29574, HD$\,$37160, HD$\,$74462, HD$\,$81192,
HD$\,105546, HD$\,148897, CD-310622, BD+521601, BD+541323).
This relation was calibrated on field stars (ATT98). The $hk$ index is similar to
the $m_1$ index, but the $v$ filter is replaced with an intermediate-band filter
centered on the Ca $H,K$ lines. 
The key advantage in using the $Ca$ filter is that the Ca $H,K$ lines, 
at fixed metal abundance, are stronger than weak metallic lines falling 
across the $v$ filter.  Therefore, the $hk$ index in the metal-poor regime 
is more sensitive to metallicity changes than the $m_1$ index. 
The second panel of Fig.~19 shows the same comparison
but with the distribution based on the ($m_0,(\bmy)_0$) calibration by ATT94
for 59 out of the 79 selected RG stars (see column 4 in Table~6). The third 
panel shows the comparison with the
metallicity distributions based on our empirical and semi-empirical
($m_0,(\vmy)_0$) MIC relations for the 79 RG stars (see columns 6 and 7 of Table~6). 
The bottom panel shows the
comparison with the metallicity distribution based on the empirical MIC relation
($m_0,(\bmy)_0$) provided by H00 for 73 out of the 79 RG stars with $-2.3<\feh<-0.4$
(see column 5 in Table~6).
This relation was calibrated using both field and cluster stars.

Data plotted in Fig.~19 indicate that spectroscopic and photometric metallicity
distributions agree quite well. The mean differences are $0.01\pm0.22$ dex (ATT98),
$0.02\pm0.22$ dex (ATT94), and $0.13\pm0.20$ dex (H00). The small difference between
current standard deviations and the original ones derived by ATT94, ATT98, and
H00  is only due to the different selection criteria. The mean differences for current
empirical and semi-empirical MIC relations are $-0.15\pm0.15$, $-0.06\pm0.18$ dex
($m_0,(\vmy)_0$) and  $-0.13\pm0.18$, $-0.10\pm0.20$ dex ($m_0,(\umy)_0$, see 
columns 8 and 9 in Table~6). Table~5 lists the mean metallicity difference for the 
entire sample according to current MIC relations and to the other calibrations adopted 
in Fig.~19. These data show that metal abundance estimates for the ATT98 
star sample based on theoretical scaled-solar MIC relations are systematically 
shifted of $\sim$ -0.24 dex towards metal-poor values, while metallicities 
estimated by adopting the empirical MIC relations are shifted of 
$\sim$ -0.14 dex, as already showed in Figs. 13 and 17. On the other hand, 
abundance estimates based on the semi-empirical MIC relations show 
a small shift $\sim$ -0.09 dex (see also Fig.~18).

\section{Conclusions and final remarks}

We provided new empirical, semi-empirical, and theoretical calibrations 
of the \strom $m_1$ metallicity index using globular cluster RG stars 
and new sets of semi-empirical and theoretical CTRs. We ended up with 
three independent sets of MIC relations to estimate iron abundance of 
cluster and field RG stars. The current MIC relations have been validated 
by adopting GCs and field RGs with known spectroscopic iron abundances in 
the Zinn \& West metallicity scale. 

Current MIC relations when compared with similar relations available in the 
literature present several key advantages:
{\em i\/}) {\em Sample selection -} In order to provide a robust calibration of the 
different MIC relations we adopted a large sample of cluster RG stars. 
{\em Bona fide\/} cluster RGs were selected using both an optical-NIR color-color 
($\umj,\,\bmh$) plane and the proper motion. Our selection was conservative to 
pin point only RG stars not affected by peculiar colors/spectra. The GCs 
(M92, M13, NGC$\,$1851, 47~Tuc) selected to calibrate the MIC relations cover a 
broad range in metallicity ($-2.2 \le \feh \le -0.7$) and are marginally 
affected by reddening uncertainties ($E(\bmv) \le 0.04$). 
{\em ii\/}) {\em Calibrations anchored to the GC metallicity scale -} The main advantage 
in using GGCs as calibrators is that their iron abundances are well-known and 
obey to a well-defined metallicity scale.  
{\em iii\/}) {\em Stronger sensitivity to effective temperature -} In order to derive 
the MIC relations we adopted the \umy\ and the \vmy\ color. These colors 
when compared with the \bmy\ color presents a stronger sensitivity to the 
effective temperature. Owing to this intrinsic properties the MIC relations
in the $m_1$ vs CI planes are linear over a large color range.  
{\em iv\/}) {\em Semi-empirical and theoretical calibration of MIC relations -} In order 
to constrain the occurrence of deceptive systematic uncertainties affecting the 
new empirical MIC relations we also derived independent semi-empirical and 
theoretical relations. Stellar isochrones covering a broad range of metal abundances 
($-2.6 \le \feh \le -0.6$) have been transformed into the observational plane 
by adopting the semi-empirical CTRs provided by CVGB04 and the theoretical 
CTRs provided by CK06. The comparison between theory and observations indicates 
that the $\alpha$-enhanced isochrones transformed using the $\alpha$-enhanced 
CTRs by CVGB04 and by CK06 account for stellar distributions in different \strom 
CMDs and in different color-color planes ($m_1$ vs CI). A similar agreement between 
theory and observations in the multiband \strom CMDs is also shown by 
scaled-solar isochrones transformed using the scaled-solar CTRs by CK06.          
On the other hand, the comparison of the same tracks in the color-color 
planes is less satisfactory. In the metal-poor regime they appear to be, 
at fixed $m_1$ value, slightly hotter, while in the metal-intermediate 
regime they agree quite well with observations. Oddly enough, in the more 
metal-rich regime the isochrones transformed using the CTRs by CK06 seem to be,
at fixed $m_1$ value, slightly cooler than observations. This evidence suggests 
that the above difference might be due to the approach adopted to fix the 
zero-points of CTRs as a function of the metal-content. This finding is only 
marginally affected by uncertainties in the evolutionary models, since we 
are using homogeneous sets of cluster isochrones based on the same theoretical 
framework.

Current MIC relations also present two main drawbacks: 
{\em i\/}) {\em Stronger sensitivity to reddening uncertainties -} The use of MIC relations 
based on the \umy\ and the \vmy\ color are more affected by uncertainties in reddening 
corrections than the MIC relations using the \bmy\ color. To partially overcome this 
thorny problem we also derived new MIC relations for the reddening-free metallicity 
index $[m]$.  
{\em ii\/}) {\em Use of the \strom $u-$band -} The use of the $u-$band data, at fixed 
photometric error, is more demanding concerning the observing time, when 
compared with the $v,b,y-$band data.

The comparison between the iron abundance estimates based on the different sets 
of MIC relations and the spectroscopic measurements available in the literature 
brings forward several interesting findings.

{\em i\/}) {\em Empirical calibration of MIC relations -} Iron abundances for RG stars 
in five GGCs (M71, NGC$\,$288, NGC$\,$362, NGC$\,$6397, NGC$\,$6752) based on the empirical 
calibration of MIC relations agree quite-well with spectroscopic measurements. 
The difference is $0.04\pm 0.03$ dex with an intrinsic dispersion $\sigma = 0.11$ dex. 
It is worth mentioning that the agreement applies not only to metal-intermediate 
clusters with low reddening corrections (NGC$\,$288, $\feh=-1.40$, $E(\bmv)=0.01$), 
but also to more metal-rich clusters with high reddening corrections 
(M71, $\feh=-0.73$, $E(\bmv)=0.31$). 
Furthermore, the difference between photometric and spectroscopic metallicities
for a sample of 79 field RG stars from the list of ATT94 and ATT98 covering a 
broad range in metal abundances ($-2.4 \le \feh \le -0.5$), is on average 
$-0.14\pm 0.01$ dex, with $\sigma = 0.17$ dex. The spread is mainly due to 
photometric errors, uncertainties in reddening corrections, and spectroscopic 
errors. 

{\em ii\/}) {\em Semi-empirical calibration of MIC relations -} Iron abundances for RG stars 
in nine GGCs (M71, NGC$\,$288, NGC$\,$362, NGC$\,$6397, NGC$\,$6752, NGC$\,$104, M92, M13, NGC$\,$1851)  
based on the semi-empirical calibration of the MIC relations agree quite-well with 
spectroscopic measurements. The difference is $0.04\pm 0.03$ dex with an intrinsic 
dispersion $\sigma = 0.10$ dex. Photometric iron abundance estimates and spectroscopic 
measurements for the sample of 79 field RG stars also agree quite-well, and indeed no 
clear trend with iron abundance is present. Moreover, the difference on average is  
$-0.09\pm 0.03$ dex with $\sigma = 0.19$ dex. 
The semi-empirical MIC relations seem to be the most robust relations
to estimate the iron abundance of field and cluster RG stars, and cover a 
very broad metallicity range i.e. from very metal-poor up to half solar iron 
abundance.
 
{\em iii\/}) {\em Theoretical calibration of MIC relations -} Iron abundances for RG stars 
in nine GGCs (M71, NGC$\,$288, NGC$\,$362, NGC$\,$6397, NGC$\,$6752, NGC$\,$104, M92, 
M13, NGC$\,$1851) based on the theoretical scaled-solar calibration of the MIC relations 
are slightly more metal-poor than spectroscopic measurements. The difference on average is 
$-0.11\pm0.03$ dex, with $\sigma=0.14$ dex.  Although, these estimates are still within 
the $1\sigma$ uncertainties, the systematic difference is clear and becomes of the 
order of 0.2 dex for a few metal-intermediate clusters. The same outcome applies 
to the comparison between photometric and spectroscopic iron abundances of the 
79 field stars by ATT94 and ATT98. The difference on average is $-0.24\pm 0.03$ dex, with 
$\sigma = 0.15$ dex. This suggests that the above difference might be due to the 
approach adopted to fix the zero-points of CTRs as a function of the metal 
content. The evidence that the scaled-solar MIC relations underestimate the 
iron abundance of both GCs and field stars requires larger samples, in particular 
of halo and disk RG stars, before we can reach firm conclusions on its nature.        

The current findings are suggesting two possible avenues for further improve the 
intrinsic accuracy of the \strom MIC relations to estimate metal abundances. 
{\em i\/}) New sets of BCs and CTRs for the \strom bands based on new atmosphere 
models are becoming available. However, we still lack firm constraints on the 
impact that the different approaches adopted to calibrate the zero-points have 
on the CTRs. 
{\em ii\/}) During the last few years the introduction of multi-object spectrographs 
such as {\sc FLAMES}  and {\sc FORS} at the ESO/VLT; {\sc HYDRA} at CTIO4m; 
{\sc AAO}mega at the AAT; {\sc GMOS} at Gemini; and {\sc DEIMOS} at Keck provided 
a wealth of new spectroscopic data for evolved stellar populations both in GCs 
(Carretta 2006, Shetrone 2003) and in Local Group dwarf galaxies 
(Coleman et al.\ 2005, Majewski et al.\ 2005). However, we still lack homogeneous 
sets of iron and heavy element abundances for sizable samples of halo and disk 
stellar populations.   
The lack of accurate \strom photometry and spectroscopic abundances is even more
severe in the metal-rich regime. In particular, current MIC calibrations need to
be extended at iron abundaces \feh $> -0.70$, but this means, at least for GCs, the
non trivial effort to collect accurate \strom photometry for GCs in the Galactic
bulge.
Moreover and even more importantly, a significant fraction of current iron 
abundances of field stars are not in the popular metallicity scale of GGCs.  
The above circumstantial evidence appears relevant steps in the use of \strom 
MIC relations to provide robust estimates of the metallicity distribution in 
complex stellar systems.   

The referee noted that calibrations based on cluster RG stars that present
either a different zero-point or a different slope than calibrations based on
field stars make the application to field RG stars questionable. This evidence
is somehow supported by the fact that field RG stars are either CN-weak or
CH-strong (Langer, Suntzeff, \& Kraft 1992), while the distribution of CN and CH
stars among RG stars differ from cluster to cluster (Gratton et al.\ 2004, and
references therein). It is clear that the cluster environment affects this
phenomenon, but no satisfying explanation has been found. However, the
application of empirical and semi-empirical calibrations to field stars for
which are available accurate spectroscopic measurements indicates that the
difference in the zero-point is at most of the order of -0.15 dex, while the
dispersion is smaller than 0.20 dex. This means that these calibrations,
within the quoted uncertainties, can be applied to estimate the metal 
abundance of field RG stars. However, the tests we performed do not allow us 
to figure out whether the systematic difference in the zero-point between 
photometric estimates and spectroscopic measurements for field stars is
intrinsic and not caused by limits in the cluster calibrated MIC relations.

\acknowledgements
It is a real pleasure to thank F. Castelli not only for sending us
bolometric corrections and color indices for \strom bands, but 
also for many useful suggestions. We also thank P.E. Nissen for  
enlightening suggestions concerning the \strom photometric system, 
M. Cignoni, S. Degl'Innocenti, P. G. Prada Moroni, and S. N. Shore 
for the simulations of the Galactic model.
We acknowledge the referee, B. Twarog, for his pertinent comments 
and suggestions that helped us to improve the content and the 
readability of the manuscript.
This work was partially supported by PRIN-INAF2005 (P.I.: A. Buzzoni),
"A laboratory to investigate stellar populations: new models and 
diagnostics", and by PRIN-INAF2004 (P.I.: M. Bellazzini),
"A hierarchical merging tale told by stars: motions, ages and chemical
compositions within structures and substructures of the Milky Way",
and by Particle Physics and Astronomy Research Council (PPARC).
This publication makes use of data products from VizieR (Ochsenbein, 
Bauer, \& Marcout 2000) and from the Two Micron All Sky Survey, which 
is a joint project of the University of Massachusetts and the Infrared 
Processing and Analysis Center/California Institute of Technology, funded 
by the National Aeronautics and Space Administration and the National 
Science Foundation.



\bibliographystyle{aa}

\begin{thebibliography}{}

\bibitem[Anthony-Twarog \& Twarog 1994]{twa94} Anthony-Twarog, B.J., \& Twarog, B.A. 1994, \aj, 107, 1577 (ATT94)
\bibitem[Anthony-Twarog, Twarog, \& Craig 1995] {} Anthony-Twarog, B.J., Twarog, B.A., \& Craig,  1995, \pasp, 107, 32
\bibitem[Anthony-Twarog \& Twarog 1998]{twa98} Anthony-Twarog, B.J., \& Twarog, B.A. 1998, \aj, 116, 1922  (ATT98)
\bibitem[Anthony-Twarog \& Twarog 2000]{twa00} Anthony-Twarog, B.J., \& Twarog, B.A. 2000, \aj, 120, 3111 (ATT00)
\bibitem[Bessell 2005]{bessel} Bessell, M. S.  2005, ARA\&A, 43, 293
\bibitem[Bond 1980]{bond80} Bond, H.E. 1980, \apjs, 44, 517
\bibitem[Bell \& Gustafsson 1978]{be78} Bell, R.A., \& Gustafsson, B. 1978, A\&AS, 34, 229 
\bibitem[Briley, Cohen, \& Stetson 2002] {} Briley, M.M., Cohen, J.G., Stetson, P.B. 2002, \apj, 579, 17
\bibitem[Briley et al.\ 2004] {} Briley, M., Harbeck, D., Smith, G.H., \& Grebel, E. 2004, \aj, 127, 1588
\bibitem[Calamida et al.\ 2005]{io05} Calamida, et al.\ 2005, AJ, 634, L69 
\bibitem[Cardelli et al.\ 1989]{card89} Cardelli, J.A., Clayton, G.C., \&  Mathis, J.S.  1989, \apj, 345, 245 
\bibitem[Carretta et al.\ 2000]{car00} Carretta, E., Gratton, R.G., Clementini, G., Fusi Pecci, F. 2000, \apj, 533, 215
\bibitem[Carretta 2006]{car06} Carretta, E. 2006, \aj, 131, 1766
\bibitem[Cassisi et al.\ 2004]{cas04} Cassisi, S., Salaris, M., Castelli, F., \& Pietrinferni, A.  2004, \apj, 616, 498  
\bibitem[Castellani et al.\ 2001] {cas04} Castellani, V., Degl'Innocenti, S., Petroni, S., Piotto, G., 2001, MNRAS, 324, 167
\bibitem[Castellani et al.\ 2002] {cas04} Castellani, V., Cignoni, M., Degl'Innocenti, S., Petroni, S., Prada Moroni, P.G. 2002, MNRAS, 334, 69
\bibitem[Castellani et al.\ 2006] {cas04} Castellani, V., Iannicola, G., Bono, G., Zoccali, M., Cassisi, S., Buonanno, R. 2006, A\&A, 446, 569
\bibitem[Castellani et al.\ 2007] {} Castellani, V., et al.\ 2007, ApJ, accepted, astro-ph/0703401  
\bibitem[Castelli \& Kurucz 2006]{CK06} Castelli, F., \& Kurucz, R. L. 2006, A\&A, 454, 333 (CK06)
\bibitem[Cignoni et al.\ 2006] {} Cignoni, M., Degl'Innocenti, S., Prada Moroni, P.G., Shore, S.N, A\&A. 459, 783
\bibitem[Clem et al.\ 2004]{clem04} Clem, J.L., VandenBerg, Don A., Grundahl, F., \& Bell, R. A. 2004, \aj, 127, 1227 (CVGB04)
\bibitem[Coleman et al.\ 2005]{col05} Coleman, M. G., Da Costa, G. S., Bland-Hawthorn, J. 2005, \aj, 130, 1065
\bibitem[Cousins 1976]{cou} Cousins, A.W.J. 1976, MNSSA, 35, 70
\bibitem[Crawford 1975]{craw75} Crawford, D. L. 1975, \aj, 80, 955
\bibitem[Crawford et al.\ 1976]{craw76} Crawford, D.L., \& Mandwewala, L. 1976, \pasp, 88, 917
\bibitem[Crawford 1979] {} Crawford, D.L. 1979, \aj, 84, 1858
\bibitem[Cudworth 1976]{cud} Cudworth, K.M. 1976,\aj, 81, 11
\bibitem[Cutri et al.\ 2003]{cutri} Cutri, R.M. \& 2MASS collaboration 2003, www.ipac.caltech.edu/2mass 
\bibitem[Faria et al.\ 2007]{faria} Faria, D., Feltzing, S. Lundstrom, I., Gilmore, G., Wahlgren, G.M., 
Ardeberg, A., Linde, P. 2007, A\&A, 465, 357
\bibitem[Feltzing \& Gilmore 2000]{felz} Feltzing, S.\& Gilmore, G. 2000, A\&A, 355, 949
\bibitem[Ferraro et al.\ 1999]{ferraro99} Ferraro, F.R., Messineo, M., Fusi Pecci, F., de Palo, M.A., Straniero, O., Chieffi, A., Limongi, M. 1999, \aj, 118, 1738
\bibitem[Freyhammer et al.\ 2005]{frey05} Freyhammer, L. M., et al.\ 2005, \apj, 623, 860
\bibitem[Gratton, Sneden \& Carretta 2004]{gra04} Gratton, R., Sneden, C. \& Carretta, E. 2004, ARA\&A, 42, 385
\bibitem[Grebel \& Richtler 1992] {} Grebel, E.K., Richtler, T. 1992, A\&A, 253, 359
\bibitem[Grundahl et al.\ 1998]{gru98} Grundahl, F., Vandenberg, Don A., Andersen, M.I. 1998, \apj, 500, 179
\bibitem[Grundahl et al.\ 1999]{gru99} Grundahl, F., Catelan, M., Landsman, W.B., Stetson, P. B., Andersen, M. I. 1999, \apj, 524, 242
\bibitem[Grundahl, Stetson, \& Andersen  2002]{gru02} Grundahl, F., Stetson, P.B., Andersen, M.I. 2002, A\&A, 395, 481
\bibitem[Haywood 2001]{hay01} Haywood, M. 2001, MNRAS, 325, 1365
\bibitem[Harbeck, Smith, \& Grebel 2003] {} Harbeck, D., Smith, G.H., Grebel, E.K. 2003, \aj, 125, 197
\bibitem[Harris 1996]{h96} Harris, W.E. 1996, \aj, 112, 1487
\bibitem[Harris 2003]{h03} Harris, W.E. 2003, Catalog of Parameters for Milky Way Globular Clusters: The Database Hamilton: McMaster Univ., http://physun.physics.mcmaster.ca/~harris/mwgc.dat 
\bibitem[Hauck \& Mermilliod 1998] {} Hauck, B., \& Mermilliod, M. 1998, A\&AS, 129, 431
\bibitem[Hilker 2000]{hilker00} Hilker, M. 2000, A\&A, 355, 994 (H00)
\bibitem[Hilker \& Richtler 2000]{hilkri00} Hilker, M., \& Richtler, T. 2000, A\&A, 362, 895 (HR00)
\bibitem[Johnson \& Morgan 1953]{johnson} Johnson, H.L. \& Morgan 1953, \apj, 117,313
\bibitem[Kraft et al.\ 1992]{Kra92}  Kraft, R.P., Sneden, C., Langer, G.E., \& Prosser, C.F. 1992, \aj, 104, 645
\bibitem[Kraft \& Ivans 2003] {} Kraft, R.P., Ivans, I. I. 2003, \pasp, 115, 143
\bibitem[Kayser et al.\ 2006]{Kay06} Kayser, A., Hilker, M., Richtler, T., \& Willemsen, P. G.  2006, A\&A, 458, 777  
\bibitem[Langer, Suntzeff, \$ Kraft 1992] {} Langer, G. E., Suntzeff, N. B.,\& Kraft, R. P 1992, \pasp, 104,523
\bibitem[Majewsk]{maj05} Majewski, S.~R., et al.\ 2005, \aj, 130, 2677
\bibitem[Nissen 1988]{nissen88} Nissen, P.E. 1988, A\&A, 199, 146
\bibitem[Nissen \& Schuster 1991]{nissen91} Nissen, P. E., \& Schuster, W. J. 1991, A\&A, 251, 457
\bibitem[Nissen 1994]{nissen94} Nissen, P.E. 1994, RMxAA, 29, 129
\bibitem[Nordstrom et al.\ 2004] {} Nordstrom, B., et al.\ 2004, A\&A, 418, 989 
\bibitem[Olsen 1983]{ols83} Olsen, E.H. 1983, A\&AS, 54, 55  
\bibitem[Olsen 1984]{ols84} Olsen, E.H. 1984, A\&AS, 57, 443
\bibitem[Olsen 1988]{ols88} Olsen, E.H. 1988, A\&A, 189, 173
\bibitem[Olsen 1993]{ols93} Olsen, E.H. 1993, A\&AS, 102, 89 (O93)
\bibitem[Olsen 1995]{ols95} Olsen, E.H. 1995, A\&A, 295, 710 (O95)
\bibitem[Percival et al.\ 2002]{perc02} Percival, S. M., Salaris, M., van Wyk, F., Kilkenny, D. 2002, \apj, 573, 174
\bibitem[Pietrinferni et al.\ 2004]{pietri04} Pietrinferni, A., Cassisi, S., Salaris, M., Castelli, F. 2004, \apj, 612, 168
\bibitem[Pietrinferni et al.\ 2006]{pietri06} Pietrinferni, A., Cassisi, S., Salaris, M. and Castelli, F. 2006, \apj, 642, 697
\bibitem[Rakos \& Schombert 2004]{rakos_01} Rakos, K., \& Schombert, J. 2004, AJ, 127, 1502 
\bibitem[Rakos \& Schombert 2005]{rakos_02} Rakos, K., \& Schombert, J. 2005, \pasp, 117, 245
\bibitem[Rey et al.\ 2004]{rey04} Rey, S-C., Lee, Y-W., Ree, C.H., Joo, J-M., Sohn, Y-J., Walker, A.R. 2004, \aj, 127, 958
\bibitem[Richter et al.\ 1999]{ric99} Richter, P., Hilker, M., \&  Richtler, T. 1999, A\&A, 350, 476 
\bibitem[Richtler 1989]{richl89} Richtler, T. 1989, A\&A, 211, 199
\bibitem[Rieke \& Lebofsky 1985]{rl85} Rieke, G.H., \& Lebofsky, M.J. 1985, \apj, 288, 618 
\bibitem[Rosenberg et al.\ 2000]{ros} Rosenberg, A., Aparicio, A., Saviane, I., \& Piotto, G. 2000, A\&AS, 145, 451 
\bibitem[Rutledge et al.\ 1997]{rutledge} Rutledge, G.A., Hesser, J.E., \& Stetson, P.B. 1997, \pasp, 109, 907
\bibitem[Salaris et al.\ 1993]{sa93} Salaris, M., Chieffi, A., \& Straniero, O. 1993, \apj, 414, 580 
\bibitem[Schlegel et al.1998]{schlegel98} Schlegel, D.J., Finkbeiner, D.P., Davis, M., 1998, \apj, 500, 525 
\bibitem[Schuster \& Nissen 1988]{shu88} Schuster, W.J., Nissen, P.E. 1988, A\&AS, 221, 65 (SN88)
\bibitem[Schuster \& Nissen 1989]{shu89} Schuster, W.J., Nissen, P.E. 1989, A\&AS, 73, 225 (SN89)
\bibitem[Shetrone 2003]{she03} Shetrone, M.D 2003, \apj, 585, L45
\bibitem[Skrutskie et al.\ 2006] {} Skrutskie, M.F., Cutri, R.M., Stiening, R., Weinberg, M.D., Schneider, S., Carpenter, J.M., et al.\ 2006, \aj, 131, 1163
\bibitem[Smith 1987]{1648} Smith, G. H. 1987, \pasp, 99, 67 
\bibitem[Smith 1988]{1649} Smith, G. H. 1988, \pasp, 100, 1104 
\bibitem[Smith, Dupree, \& Churchill 1992]{smith92} Smith, G.H., Dupree, A. K., \& Churchill, C. W. 1992, \aj, 104, 2005 
\bibitem[Stanford et al.\ 2004] {} Stanford, L.M., Da Costa, G.S., Norris, J.E., Cannon, R.D. 2004, MemSAIt, 75, 290 
\bibitem[Stetson 1987]{ste87} Stetson, P.B. 1987, \pasp, 99, 191 
\bibitem[Stetson 1991]{ste91} Stetson, P.B. 1991, \aj, 102, 589 
\bibitem[Stetson 1994]{ste94} Stetson, P.B. 1994, \pasp, 106, 250 
\bibitem[Str\"omgren 1966]{strom66} Str\"omgren, B. 1966, Ann.Rev. A\&A, 4, 433
\bibitem[Twarog \& Anthony-Twarog 1991] {} Twarog, B.A., Anthony-Twarog, B.J. 1991, \aj, 101, 237
\bibitem[Twarog et al.\ 1997] {} Twarog, B.A., Keith M., Anthony-Twarog, B.J. 1997, \aj, 114, 2556
\bibitem[Zacharias et al.\ 2004]{1667} Zacharias et al., 2004, \apj, 127, 3043
\bibitem[Zinn 1985]{zinn85} Zinn, R. 1985, \apj, 293, 424
\bibitem[Zinn \& West 1984]{zinn} Zinn, R., \& West, M. J. 1984, \apjs, 55, 45

\end{thebibliography}


\begin{deluxetable}{lccc}
\tablewidth{0pt}
\tablecaption{Log of observations.\label{tbl-1}}
\tablehead{
\colhead{Cluster}&
\colhead{Telescope\tablenotemark{a}}&
\colhead{Date\tablenotemark{b}}&
\colhead{Reference\tablenotemark{c}}\\
\colhead{(1)}&
\colhead{(2)}&
\colhead{(3)}&
\colhead{(4)}}
\startdata
           \multicolumn{4}{c}{}  \\  
NGC$\,$6341 (M92)& NOT & June 27/July 2, 1995 & 1\\ 
NGC$\,$6397& Danish & May, 1997     & 1 \\
NGC$\,$6205 (M13)& NOT    & June 27/July 2, 1995 & 2 \\
NGC$\,$6752& Danish & October, 1997 & 1 \\
NGC$\,$288 & Danish & October, 1997 & 1 \\
NGC$\,$1851& Danish & October, 1997 & 1 \\
NGC$\,$362 & Danish & October, 1997 & 1 \\
NGC$\,$6838 (M71)& NOT & June 26/July 2, 1995 & 3 \\
NGC$\,$104 (47~Tuc)& Danish & October, 1997 & 3 \\
\enddata			     
\tablenotetext{a}{Observations were collected with the Nordic Optical Telescope (NOT)
operated at La Palma and the 1.54m Danish Telescope operated at ESO (La Silla).}
\tablenotetext{b}{Date of observations.}
\tablenotetext{c}{(1) Grundahl et al.\ (1999); (2) Grundahl et al.\ (1998);
(3) Grundahl, Stetson \& Andersen\ (2002).}
\end{deluxetable}				   


\begin{deluxetable}{lccc}
\tablewidth{0pt}
\tablecaption{Sample of globular clusters adopted to calibrate and to 
validate the \strom metallicity index.\label{tbl-2}}
\tablehead{
\colhead{ID}&
\colhead{$\feh$\tablenotemark{a}}&
\colhead{$\sigma \feh$\tablenotemark{b}}&
\colhead{$E(\bmv)$\tablenotemark{c}}\\
\colhead{(1)}&
\colhead{(2)}&
\colhead{(3)}&
\colhead{(4)}}
\startdata
           \multicolumn{4}{c}{}  \\  
M92     & -2.24\tablenotemark{d} & 0.10 & 0.02 \\ 
NGC$\,$6397& -1.91 & 0.14 & 0.18 \\
M13     & -1.65 & 0.06 & 0.02 \\
NGC$\,$6752& -1.54 & 0.09 & 0.04 \\
NGC$\,$288 & -1.40 & 0.12 & 0.01\tablenotemark{e} \\
NGC$\,$1851& -1.33 & 0.10 & 0.02 \\
NGC$\,$362 & -1.27 & 0.07 & 0.05 \\
M71     & -0.73\tablenotemark{f} & 0.05 & 0.31\tablenotemark{g} \\
NGC$\,$104 & -0.71 & 0.05 & 0.04 \\
\enddata			     
\tablenotetext{a}{Cluster metal abundances according to Rutledge et al.\ (1997)
in the metallicity scale by Zinn \& West (1984) and Zinn (1985).}  
\tablenotetext{b} {Errors on metal abundances according to 
Zinn \& West (1984).}
\tablenotetext{c} {Values from the GC catalog by Harris (1996,2003).}
\tablenotetext{d} {Value from Zinn \& West (1984).} 
\tablenotetext{e} {Value from Schlegel et al.\ (1998), and Calamida et al.\ (2005).} 
\tablenotetext{f} {Value from GC catalog by Harris (1996,2003).} 
\tablenotetext{g} {Value from Schlegel et al.\ (1998).} 
\end{deluxetable}				   


\begin{deluxetable}{lccccc}
\tablewidth{0pt}
\tabletypesize{\scriptsize}
\tablecaption{Multilinear regression coefficients for the \strom  
metallicity index: $m = \alpha + \beta\cdot\feh + \gamma\cdot CI +
\delta\cdot(\feh\cdot CI)$.\label{tbl-3}}
\tablehead{
\colhead{Relation}&
\colhead{$\alpha$}&
\colhead{$\beta$}&
\colhead{$\gamma$}&
\colhead{$\delta$}&
\colhead{Multi Correlation}\\
\colhead{(1)}&
\colhead{(2)}&
\colhead{(3)}&
\colhead{(4)}&
\colhead{(5)}&
\colhead{(6)}}
\startdata
\multicolumn{6}{c}{Empirical based on selected GCs}  \\  
$m_0, (\umy)_0$ & -0.374 & -0.132$\pm$0.0014 & 0.300$\pm$0.0005 & 0.098$\pm$0.0003 & 0.99 \\    
$m_0, (\vmy)_0$ & -0.312 & -0.096$\pm$0.0015 & 0.513$\pm$0.001  & 0.154$\pm$0.0006 & 1.00 \\    
$[m], (\umy)_0$ & -0.340 & -0.130$\pm$0.0014 & 0.348$\pm$0.0005 & 0.093$\pm$0.0003 & 1.00 \\    
$[m], (\vmy)_0$ & -0.278 & -0.092$\pm$0.0015 & 0.592$\pm$0.001  & 0.136$\pm$0.0006 & 1.00 \\    

\multicolumn{6}{c}{Theoretical based on transformations by CK06}  \\  
$m_0, (\umy)_0$ & -0.196 & -0.043$\pm$0.0009 & 0.232$\pm$0.0004 & 0.061$\pm$0.0003 & 1.00\\    
$m_0, (\vmy)_0$ & -0.169 & -0.025$\pm$0.0008 & 0.401$\pm$0.0007 & 0.094$\pm$0.0004 & 1.00\\    
$[m], (\umy)_0$ & -0.184 & -0.048$\pm$0.0009 & 0.287$\pm$0.0004 & 0.058$\pm$0.0003 & 1.00\\    
$[m], (\vmy)_0$ & -0.145 & -0.022$\pm$0.0008 & 0.491$\pm$0.0007 & 0.080$\pm$0.0004 & 1.00\\    

\multicolumn{6}{c}{Semi-empirical based on transformations by CVGB04} \\  
$m_0, (\umy)_0$ & -0.323 & -0.099$\pm$0.005 & 0.294$\pm$0.002 & 0.094$\pm$0.001 & 1.00 \\    
$m_0, (\vmy)_0$ & -0.309 & -0.090$\pm$0.002 & 0.521$\pm$0.001 & 0.159$\pm$0.001 & 0.99 \\    
$[m], (\umy)_0$ & -0.289 & -0.094$\pm$0.005 & 0.337$\pm$0.002 & 0.084$\pm$0.001 & 1.00 \\    
$[m], (\vmy)_0$ & -0.251 & -0.070$\pm$0.005 & 0.585$\pm$0.004 & 0.131$\pm$0.002 & 1.00 \\    
\enddata

\end{deluxetable}				   

\begin{deluxetable}{lccccccccc}
\tabletypesize{\scriptsize}
\tablewidth{0pt}
\tablecaption{Metallicity estimates based on the empirical, the theoretical and the
semi-empirical MIC relations of Table 3 for the nine globular clusters we adopted to calibrate
and validate the relations.\label{tbl-4}}
\tablehead{
\colhead{Relation}&
\colhead{M92}&
\colhead{NGC6397}&
\colhead{M13}&
\colhead{NGC6752}&
\colhead{NGC288}&
\colhead{NGC1851}&
\colhead{NGC362}&
\colhead{M71}&
\colhead{NGC104}\\
\colhead{(1)}&
\colhead{(2)}&
\colhead{(3)}&
\colhead{(4)}&
\colhead{(5)}&
\colhead{(6)}&
\colhead{(7)}&
\colhead{(8)}&
\colhead{(9)}&
\colhead{(10)}}
\startdata
\multicolumn{10}{c}{Empirical based on selected GCs}  \\  
$m_0,(\umy)_0$ & $\ldots$ & -1.97$\pm$0.24 & $\ldots$ & -1.56$\pm$0.19 & -1.19$\pm$0.15 & $\ldots$ & -1.21$\pm$0.25 & -0.62$\pm$0.25 & $\ldots$ \\    
$m_0,(\vmy)_0$ & $\ldots$ & -2.00$\pm$0.20 & $\ldots$ & -1.66$\pm$0.18 & -1.32$\pm$0.11 & $\ldots$ & -1.26$\pm$0.23 & -0.60$\pm$0.24 & $\ldots$ \\    
$[m],(\umy)_0$ & $\ldots$ & -1.93$\pm$0.24 & $\ldots$ & -1.55$\pm$0.21 & -1.20$\pm$0.16 & $\ldots$ & -1.23$\pm$0.26 & -0.65$\pm$0.22 & $\ldots$ \\    
$[m],(\vmy)_0$ & $\ldots$ & -1.97$\pm$0.18 & $\ldots$ & -1.65$\pm$0.20 & -1.30$\pm$0.11 & $\ldots$ & -1.23$\pm$0.23 & -0.50$\pm$0.26 & $\ldots$ \\    
\multicolumn{10}{c}{Theoretical based on transformations by CK06}  \\  
$m_0,(\umy)_0$ & -2.33$\pm$0.14 & -2.13$\pm$0.16 & -1.91$\pm$0.12 & -1.76$\pm$0.18 & -1.29$\pm$0.12 & -1.56$\pm$0.21 & -1.42$\pm$0.29 & -0.86$\pm$0.20 & -0.73$\pm$0.26 \\	  
$m_0,(\vmy)_0$ & -2.30$\pm$0.14 & -2.12$\pm$0.17 & -1.90$\pm$0.10 & -1.77$\pm$0.18 & -1.29$\pm$0.11 & -1.55$\pm$0.23 & -1.35$\pm$0.26 & -0.67$\pm$0.25 & -0.62$\pm$0.29 \\	  
$[m],(\umy)_0$ & -2.36$\pm$0.12 & -2.14$\pm$0.19 & -1.91$\pm$0.15 & -1.73$\pm$0.18 & -1.27$\pm$0.13 & -1.56$\pm$0.21 & -1.42$\pm$0.28 & -0.87$\pm$0.18 & -0.74$\pm$0.24 \\	  
$[m],(\vmy)_0$ & -2.29$\pm$0.14 & -2.06$\pm$0.17 & -1.89$\pm$0.10 & -1.76$\pm$0.18 & -1.28$\pm$0.11 & -1.55$\pm$0.23 & -1.34$\pm$0.26 & -0.61$\pm$0.27 & -0.61$\pm$0.29 \\	  
\multicolumn{10}{c}{Semi-empirical based on transformations by CVGB04} \\  
$m_0,(\umy)_0$ & -2.02$\pm$0.16 & -1.85$\pm$0.18 & -1.69$\pm$0.11 & -1.56$\pm$0.14 & -1.27$\pm$0.10 & -1.44$\pm$0.15 & -1.28$\pm$0.20 & -0.75$\pm$0.20 & -0.77$\pm$0.18 \\    
$m_0,(\vmy)_0$ & -1.99$\pm$0.17 & -1.83$\pm$0.18 & -1.67$\pm$0.10 & -1.56$\pm$0.14 & -1.26$\pm$0.10 & -1.41$\pm$0.16 & -1.21$\pm$0.19 & -0.64$\pm$0.20 & -0.69$\pm$0.22 \\    
$[m],(\umy)_0$ & -2.07$\pm$0.14 & -1.81$\pm$0.22 & -1.71$\pm$0.14 & -1.54$\pm$0.16 & -1.26$\pm$0.12 & -1.45$\pm$0.16 & -1.29$\pm$0.21 & -0.75$\pm$0.19 & -0.78$\pm$0.18 \\    
$[m],(\vmy)_0$ & -1.98$\pm$0.16 & -1.79$\pm$0.19 & -1.67$\pm$0.10 & -1.56$\pm$0.14 & -1.28$\pm$0.10 & -1.42$\pm$0.16 & -1.21$\pm$0.19 & -0.59$\pm$0.21 & -0.69$\pm$0.20 \\    
\enddata			     
\end{deluxetable}				   

  
\begin{deluxetable}{lc}
\tablewidth{0pt}
\tablecaption{The difference between metallicities estimated by adopting different MIC 
relations and the spectroscopic measurements for field RG stars collected by ATT94 
and ATT98.\label{tbl-5}}
\tablehead{
\colhead{Relation}&
\colhead{ATT94/ATT98}\\
\colhead{(1)}&
\colhead{(2)}}
\startdata
           \multicolumn{2}{c}{Empirical based on selected GCs\tablenotemark{a}}  \\  
$m_0,(\umy)_0$ & -0.13$\pm$0.18 \\ 
$m_0,(\vmy)_0$ & -0.15$\pm$0.15 \\ 
$[m],(\umy)_0$ & -0.15$\pm$0.21 \\ 
$[m],(\vmy)_0$ & -0.13$\pm$0.15 \\ 
\multicolumn{2}{c}{Theoretical based on transformations by CK06\tablenotemark{a}} \\
$m_0,(\umy)_0$ & -0.26$\pm$0.15 \\ 
$m_0,(\vmy)_0$ & -0.22$\pm$0.14 \\ 
$[m],(\umy)_0$ & -0.27$\pm$0.18 \\ 
$[m],(\vmy)_0$ & -0.21$\pm$0.15 \\ 
\multicolumn{2}{c}{Semi-empirical based on transformations by CVGB04\tablenotemark{a}} \\
$m_0,(\umy)_0$ & -0.10 $\pm$0.20 \\ 
$m_0,(\vmy)_0$ & -0.06 $\pm$0.18 \\ 
$[m],(\umy)_0$ & -0.13 $\pm$0.22 \\ 
$[m],(\vmy)_0$ & -0.05 $\pm$0.18 \\ 
\multicolumn{2}{c}{Other calibrations} \\
$hk_0,(\bmy)_0$\tablenotemark{b} &  0.01$\pm$0.22 \\ 
$m_0,(\bmy)_0$\tablenotemark{c}  &  0.02$\pm$0.22 \\ 
$m_0,(\bmy)_0$\tablenotemark{d}  &  0.13$\pm$0.20 \\ 
\enddata			     
\tablenotetext{a}{The MIC relations were applied to a sample of 79 field RG stars.}  
\tablenotetext{b}{ATT98 MIC relation applied to 70 field RG stars.}  
\tablenotetext{c}{ATT94 MIC relation applied to 59 field RG stars.}
\tablenotetext{d}{H00 MIC relation applied to 73 field RG stars.}
\end{deluxetable}				   

  

\begin{deluxetable}{lcccccccc}
\tablewidth{0pt}
\tabletypesize{\scriptsize}
\tablecaption{Spectroscopic measurements and photometric estimates of iron 
abundances for the sample of 85 field RG stars collected by ATT94 and ATT98.}\label{tbl-6}
\tablehead{
\colhead{Star ID}&
\colhead{$[Fe/H]$\tablenotemark{a}} &
\colhead{$[Fe/H]$\tablenotemark{b}} &
\colhead{$[Fe/H]$\tablenotemark{c}} &
\colhead{$[Fe/H]$\tablenotemark{d}} &
\colhead{$[Fe/H]$\tablenotemark{e}} &
\colhead{$[Fe/H]$\tablenotemark{f}} &
\colhead{$[Fe/H]$\tablenotemark{g}} &
\colhead{$[Fe/H]$\tablenotemark{h}} \\
\colhead{} &
\colhead{Spec} &
\colhead{ATT98} &
\colhead{ATT94} &
\colhead{H00} &
\colhead{$m_1, vy$} &
\colhead{$m_1, uy$} &
\colhead{$m_1, vy_{sem}$} &
\colhead{$m_1, uy_{sem}$} }
\startdata
HD 97       & -1.22 & -1.31    & -1.38    & -1.02 & -1.37 & -1.31 & -1.30 & -1.37\\
HD 2665     & -1.91 & -2.14    & -1.99    & -1.59 & -2.07 & -2.01 & -1.89 & -1.89\\
HD 3008     & -1.90 & -1.80    & -1.43    & -1.83 & -1.94 & -1.94 & -1.86 & -1.92\\
HD 5426     & -2.35 & -2.22    & -2.33    & -1.76 & -2.34 & -2.26 & -2.12 & -2.07\\
HD 6268     & -2.37 & -2.38    & -2.25    & -1.95 & -2.34 & -2.35 & -2.17 & -2.21\\
HD 6755     & -1.62 & -1.85    & -1.72    & -1.31 & -1.76 & -1.60 & -1.62 & -1.59\\
HD 6833     & -1.06 & -0.67    & -1.00    & -0.83 & -1.12 & -1.19 & -1.10 & -1.23\\
HD 7595     & -0.85 & -0.56    & $\ldots$ & -0.69 & -1.07 & -1.12 & -1.06 & -1.15\\
HD 8724     & -1.76 & -1.93    & -2.01    & -1.95 & -2.22 & -2.14 & -2.09 & -2.06\\
HD 18907    & -0.83 & -0.56    & $\ldots$ & -0.42 & -0.82 & -0.69 & -0.83 & -0.86\\
HD 21581    & -1.72 & -1.68    & -1.65    & -1.43 & -1.79 & -1.69 & -1.68 & -1.66\\
HD 23798    & -2.22 & -2.08    & -1.90    & -1.98 & -2.16 & -2.15 & -2.05 & -2.09\\
HD 24616    & -0.82 & -0.13    & $\ldots$ & -0.57 & -0.95 & -0.91 & -0.94 & -1.04\\
HD 26169    & -2.31 & -2.23    & $\ldots$ & -1.64 & -2.25 & -2.21 & -2.01 & -2.00\\
HD 26297    & -1.76 & -1.75    & -1.67    & -1.79 & -1.96 & -1.96 & -1.88 & -1.92\\
HD 35179    & -0.59 & -0.79    & $\ldots$ & -0.64 & -1.01 & -0.97 & -1.00 & -1.06\\
HD 36702    & -2.03 & -2.06    & -2.06    & -2.07 & -2.17 & -2.14 & -2.08 & -2.10\\
HD 37160    & -0.53 & $\ldots$ & -0.63    & -0.56 & -0.95 & -1.09 & -0.95 & -1.16\\
HD 37828    & -1.32 & -0.67    & $\ldots$ & -1.08 & -1.34 & -1.38 & -1.31 & -1.40\\
HD 44007    & -1.61 & -1.10    & -1.23    & -1.10 & -1.43 & -1.40 & -1.36 & -1.43\\
HD 45282    & -1.51 & $\ldots$ & -1.80    & -1.10 & -1.53 & -1.29 & -1.42 & -1.37\\
HD 55496$\tablenotemark{i}$    & -1.55 & -1.00    & $\ldots$ & -0.87 & -1.21 & -1.21 & -1.18 & -1.25\\
HD 74462    & -1.53 & $\ldots$ & -1.60    & -1.46 & -1.74 & -1.71 & -1.65 & -1.69\\
HD 81192    & -0.74 & $\ldots$ & -0.82    & -0.72 & -1.09 & -1.19 & -1.07 & -1.24\\
HD 81223    & -0.79 & -0.59    & $\ldots$ & -0.75 & -1.09 & -1.09 & -1.07 & -1.16\\
HD 83212    & -1.48 & -1.46    & -1.45    & -1.47 & -1.70 & -1.75 & -1.63 & -1.73\\
HD 84903$\tablenotemark{l}$    & -2.28 & -2.86    & -2.55    & -2.47 & -2.70 & -2.63 & -2.54 & -2.51\\
HD 85773    & -2.28 & -2.50    & -2.22    & -2.24 & -2.40 & -2.35 & -2.28 & -2.28\\
HD 87140    & -1.85 & -1.77    & $\ldots$ & -1.34 & -1.80 & -1.60 & -1.66 & -1.58\\
HD 99978    & -1.03 & -0.60    & $\ldots$ & -0.78 & -1.13 & -0.97 & -1.11 & -1.06\\
HD 101063   & -1.15 & -1.06    & $\ldots$ & -0.91 & -1.25 & -1.02 & -1.20 & -1.16\\
HD 103295   & -0.98 & -1.07    & $\ldots$ & -0.87 & -1.20 & -1.20 & -1.16 & -1.28\\
HD 103545   & -2.09 & -2.33    & -2.42    & -1.98 & -2.40 & -2.35 & -2.22 & -2.20\\
HD 104893   & -1.92 & -2.08    & -1.78    & -1.92 & -2.06 & -2.07 & -1.97 & -2.02\\
HD 105546   & -1.44 & $\ldots$ & -1.33    & -0.77 & -1.13 & -1.47 & -1.08 & -1.49\\
HD 108317   & -2.34 & -2.33    & -2.48    & -1.63 & -2.40 & -2.39 & -2.11 & -2.07\\
HD 110184   & -2.31 & -2.39    & -2.18    & -2.28 & -2.40 & -2.34 & -2.28 & -2.28\\
HD 111721   & -1.31 & -1.28    & $\ldots$ & -1.22 & -1.58 & -1.44 & -1.49 & -1.47\\
HD 117220   & -0.76 & -0.93    & $\ldots$ & -0.73 & -1.08 & -1.01 & -1.06 & -1.12\\
HD 118055   & -1.75 & -1.77    & $\ldots$ & -1.80 & -1.91 & -1.89 & -1.84 & -1.86\\
HD 122956   & -1.74 & -1.66    & -1.71    & -1.67 & -1.93 & -1.92 & -1.83 & -1.86\\
HD 126238   & -1.69 & -1.74    & -1.68    & -1.43 & -1.81 & -1.77 & -1.69 & -1.72\\
HD 128188   & -1.29 & -1.51    & $\ldots$ & -1.30 & -1.57 & -1.50 & -1.50 & -1.51\\
HD 128279   & -2.18 & -2.03    & -1.97    & -1.40 & -2.05 & -1.87 & -1.81 & -1.73\\
HD 135148$\tablenotemark{i}$    & -1.88 & -1.64    & -0.99    & -1.40 & -1.57 & -1.52 & -1.53 & -1.52\\
HD 136316   & -1.85 & -1.82    & -1.54    & -1.69 & -1.85 & -1.88 & -1.77 & -1.85\\
HD 141531   & -1.61 & -1.62    & -1.57    & -1.70 & -1.87 & -1.87 & -1.79 & -1.84\\
HD 148897   & -1.16 & $\ldots$ & -0.98    & -1.10 & -1.36 & -1.42 & -1.33 & -1.43\\
HD 165195   & -2.25 & -2.27    & -2.16    & -2.29 & -2.39 & -2.35 & -2.28 & -2.28\\
HD 171496   & -1.03 & -1.33    & -1.16    & -0.78 & -0.96 & -1.15 & -0.95 & -1.22\\
HD 175305   & -1.45 & -1.38    & -1.39    & -1.15 & -1.53 & -1.38 & -1.44 & -1.43\\
HD 175329   & -0.60 & -0.34    & $\ldots$ & -0.65 & -1.05 & -1.04 & -1.05 & -1.07\\
HD 184711   & -2.31 & -2.46    & -2.46    & -2.41 & -2.50 & -2.42 & -2.38 & -2.36\\
HD 187111   & -1.95 & -1.73    & -1.65    & -1.81 & -1.95 & -1.94 & -1.87 & -1.91\\
HD 190287   & -1.37 & -1.32    & -1.09    & -0.95 & -1.30 & -1.09 & -1.24 & -1.21\\
HD 204543   & -1.78 & -1.83    & -1.85    & -1.68 & -1.98 & -2.04 & -1.87 & -1.97\\
HD 206739   & -1.60 & -1.61    & -1.57    & -1.52 & -1.80 & -1.79 & -1.71 & -1.75\\
HD 216143   & -2.13 & -2.21    & -2.01    & -1.98 & -2.23 & -2.20 & -2.10 & -2.13\\
HD 218857   & -1.87 & -2.12    & -2.15    & -1.59 & -2.16 & -2.09 & -1.95 & -1.92\\
HD 220662   & -1.60 & -1.80    & -1.75    & -1.70 & -1.93 & -1.95 & -1.83 & -1.90\\
HD 220838   & -1.72 & -1.66    & -1.72    & -1.70 & -1.86 & -1.86 & -1.79 & -1.83\\
HD 221170   & -2.01 & -2.13    & -2.01    & -2.14 & -2.36 & -2.32 & -2.23 & -2.23\\
HD 222434   & -1.94 & -1.73    & -1.56    & -1.66 & -1.86 & -1.86 & -1.78 & -1.83\\
BD +01 2916 & -1.78 & -1.81    & -1.45    & -1.95 & -2.02 & -1.98 & -1.94 & -1.96\\
BD +03 2782 & -1.94 & -1.95    & $\ldots$ & -1.95 & -2.08 & -2.12 & -1.96 & -2.04\\
BD +04 2466$\tablenotemark{i}$  & -2.07 & -1.85    & -0.85    & -0.83 & -1.17 & -0.86 & -1.14 & -1.02\\
BD +06 0648 & -1.99 & -2.12    & -1.82    & -1.99 & -2.11 & -2.06 & -2.01 & -2.02\\
BD +08 2856 & -2.00 & $\ldots$ & -1.98    & -1.93 & -2.22 & -2.29 & -2.09 & -2.18\\
BD +09 2870 & -2.33 & -2.42    & -2.37    & -2.12 & -2.46 & -2.42 & -2.29 & -2.29\\
BD +10 2495 & -1.78 & -1.96    & -2.14    & -1.55 & -2.06 & -2.02 & -1.88 & -1.89\\
BD +30 2611 & -1.36 & -1.52    & -1.25    & -1.43 & -1.61 & -1.60 & -1.56 & -1.60\\
BD +52 1601 & -1.36 & $\ldots$ & -1.49    & -1.18 & -1.54 & -1.66 & -1.46 & -1.64\\
BD +54 1323 & -1.65 & $\ldots$ & -1.85    & -1.15 & -1.63 & -1.49 & -1.93 & -1.83\\
BD -01 1792 & -0.98 & -0.54    & $\ldots$ & -0.68 & -1.02 & -0.91 & -1.00 & -1.05\\
BD -01 2582$\tablenotemark{i}$  & -2.25 & -2.07    & -1.54    & -1.28 & -1.71 & -1.31 & -1.59 & -1.38\\
BD -09 5831 & -1.80 & -1.78    & -1.87    & -1.78 & -2.06 & -2.07 & -1.94 & -2.00\\
BD -10 0548 & -1.09 & -1.66    & -1.71    & -0.95 & -1.28 & -1.18 & -1.22 & -1.28\\
BD -14 5890 & -1.99 & -2.08    & -2.01    & -1.69 & -2.13 & -2.06 & -1.96 & -1.94\\
BD -18 0271 & -2.20 & -2.25    & -1.98    & -2.22 & -2.32 & -2.27 & -2.21 & -2.22\\
BD -18 2065 & -0.67 & -0.58    & $\ldots$ & -0.56 & -0.93 & -1.00 & -0.93 & -1.08\\
CD -30 1121 & -1.59 & -1.99    & -1.82    & -1.40 & -1.82 & -1.73 & -1.69 & -1.68\\
CD -62 1346$\tablenotemark{i}$ & -1.46 & -1.35    & $\ldots$ & -0.59 & -0.92 & -0.93 & -0.91 & -1.09\\
CP -57 0680 & -0.60 & -0.57    & $\ldots$ & -0.86 & -1.21 & -1.26 & -1.18 & -1.30\\
ANON        & -1.58 & -1.57    & -1.49    & -1.47 & -1.72 & -1.80 & -1.64 & -1.77\\
TY VIR      & -1.58 & $\ldots$ & -0.98    & -1.65 & -1.75 & -1.75 & -1.70 & -1.74\\
\enddata 						     				     

\tablenotetext{a}{Spectroscopic iron abundances collected and transformed
to the Zinn \& West metallicity scale by ATT98.} 				     				     
\tablenotetext{b}{Photometric iron abundances estimated by ATT98 
adopting their $hk_0, (\bmy)_0$ metallicity calibration.} 
\tablenotetext{c}{Photometric iron abundances estimated by ATT94 
adopting their $m_0, (\bmy)_0$ metallicity calibration.} 
\tablenotetext{d}{Photometric iron abundances estimated 
adopting H00 $m_0, (\bmy)_0$ metallicity calibration.} 
\tablenotetext{e}{Photometric iron abundances estimated
adopting our $m_0, (\vmy)_0$ empirical MIC relations.} 
\tablenotetext{f}{Photometric iron abundances estimated
adopting our $m_0, (\umy)_0$ empirical MIC relations.} 
\tablenotetext{g}{Photometric iron abundances estimated
adopting our $m_0, (\vmy)_0$ semi-empirical MIC relations.} 
\tablenotetext{h}{Photometric iron abundances estimated
adopting our $m_0, (\umy)_0$ semi-empirical MIC relations.} 
\tablenotetext{i}{ $CH$-strong star according to ATT94 and ATT98.} 
\tablenotetext{l}{ Peculiar star according to Smith et al.\ (1992) and ATT98.} 
\end{deluxetable}

  
\end{document}